\documentclass[11pt]{emulateapj}  	
\newcommand{\p}{\partial}

\usepackage{amsmath}
\usepackage{natbib}
\usepackage{verbatim}

\begin{document}
\title{Dynamics of an Alfv\'en surface in core collapse supernovae}

\author{J\'er\^ome Guilet, Thierry Foglizzo \& S\'ebastien Fromang}
\affil{Laboratoire AIM, CEA/DSM-CNRS-Universit\'e Paris Diderot, IRFU/Service d'Astrophysique, \\
CEA-Saclay F-91191 Gif-sur-Yvette, France. }
\email{jerome.guilet@cea.fr}

\begin{abstract}
We investigate the dynamics of an Alfv\'en surface (where the Alfv\'en speed equals the advection velocity) in the context of core collapse supernovae during the phase of accretion on the proto-neutron star. Such a  surface should exist even for weak magnetic fields because the advection velocity decreases to zero at the center of the collapsing core. In this decelerated flow, Alfv\'en waves created by the standing accretion shock instability (SASI) or convection accumulate and amplify while approaching the Alfv\'en surface. 

We study this amplification using one dimensional MHD simulations with explicit physical dissipation (resistivity and viscosity). In the linear regime, the amplification continues until the Alfv\'en wavelength becomes as small as the dissipative scale. A pressure feedback that increases the pressure in the upstream flow is created via a non linear coupling. We derive analytic formulae for the maximum amplification and the non linear coupling  and check them with numerical simulations to a very good accuracy. Interestingly, these quantities diverge if the dissipation is decreased to zero, scaling like the square root of the Reynolds number, suggesting large effects in weakly dissipative flows. We also characterize the non linear saturation of this amplification when compression effects become important, leading to either a change of the velocity gradient, or a steepening of the Alfv\'en wave. 

Applying these results to core collapse supernovae shows that the amplification can be fast enough to affect the dynamics, if the magnetic field is strong enough for the Alfv\'en surface to lie in the region of strong velocity gradient just above the neutrinosphere. This requires the presence of a strong magnetic field in the progenitor star, which would correspond to the formation of a magnetar under the assumption of magnetic flux conservation.  An extrapolation of our analytic formula (taking into account the nonlinear saturation) suggests that the Alfv\'en wave could reach an amplitude of $B\sim 10^{15}\,{\rm G}$, and that the pressure feedback could significantly contribute to the pressure below the shock.
\end{abstract}


\section{Introduction}
The theoretical mechanism responsible for the explosion of massive stars has become a multidimensional puzzle when spherical models were ruled out by accurate numerical simulations \citep{liebendorfer01}.
In all mechanisms proposed recently, multidimensional fluid motions are essential for successful explosions of massive stars via core collapse \citep{marek09, burrows06, burrows07b}. These motions are induced by several hydrodynamic instabilities: convection in the proto-neutron star \citep{dessart06}, convection in the neutrino heated region between the neutrinosphere and the shock \citep{herant92} and the standing accretion shock instability (SASI, e.g.  \cite{blondin03,foglizzo07}). 

The effect of a magnetic field on the dynamics has been studied mainly in the context of very rapid rotation. In this case the magnetic field can be amplified to a dynamically significant strength by the winding of the initial field and by the magneto-rotational instability \citep{akiyama03, obergaulinger09}. This magnetic field could then extract enough rotational energy to power an explosion driven by magnetic jets \citep{bisnovatyi-Kogan76, shibata06, burrows07b}. 

For moderate rotation rates, the magnetic field is usually neglected because the magnetic pressure is thought to be much smaller than the thermal pressure. The criterion determining the significance of the magnetic field effects, however, may be more complex. For example, \cite{guilet10a} suggested that a magnetic field can affect SASI significantly if the Alfv\'en speed is comparable with the advection velocity, which is a much less stringent criterion in the very subsonic part of the flow. This may explain the surprising stabilization of SASI by a radial magnetic field which pressure is negligibly small ($P_{\rm mag}/P < 10^{-3}$), in the model 2DB14Am of \cite{endeve10}. 
Moreover, \cite{endeve10} showed that even in the absence of rotation, the magnetic field could be amplified by the turbulent flow driven by SASI. 
This motivates our study of the effect of a moderate magnetic field in the context of negligible rotation, when the Alfv\'en and the fluid velocities are comparable. 

The Alfv\'en surface is defined as the surface where the flow velocity equals the Alfv\'en velocity.  Such a surface exists in the context of a wind or a jet ejected from a magnetized object (e.g. the Sun) because the fluid accelerates from a subAlfv\'enic to a superAlfv\'enic speed. Symmetrically in the context of accretion of a magnetized fluid onto a solid object, if the flow is superAlfv\'enic far from the accretor it has to decelerate to a subAlfv\'enic speed before settling down onto the surface. The fundamental difference between these two situations is that an Alfv\'en wave that is propagating against the flow accumulates at the Alfv\'en surface in a decelerated flow (as its velocity is decreasing to zero), while it diverges from it in an accelerated flow. This is similar to the hydrodynamical situation at the sonic point:  in a decelerated transsonic flow, the accumulation of acoustic waves leads to the formation of a shock wave, while no such thing happens in an accelerated flow. The formation of a transAlfv\'enic discontinuity was shown to be non evolutionary by \cite{syrovatskii59}: in the presence of an arbitrarily small perturbation, it instantaneously disintegrates into several other discontinuities. What happens then at the Alfv\'en surface? To our knowledge, the only study dealing with a decelerated transAlfv\'enic flow is that of \cite{williams75}. He showed that Alfv\'en waves not only accumulate but are actually amplified while approaching the Alfv\'en surface in the sense that their energy flux increases. He proposed that this could cause a broad turbulent zone below the Alfv\'en surface.

Such an Alfv\'en surface is expected to exist in the context of core collapse supernovae during the phase of accretion onto the proto-neutron star. This holds even for a very weak magnetic field (and thus a very slow Alfv\'en speed) because the infall velocity vanishes at the center of the star. Furthermore strong Alfv\'en waves are created by SASI \citep{guilet10a} and by the convection inside the proto-neutron star \citep{suzuki08}. The primary goal of this paper is to study the evolution of these waves properties as they reach the Alfv\'en surface in the conditions prevailing during the collapse of a massive star core. \cite{williams75} made a linear analysis of high frequency Alfv\'en waves in a cold supersonic and inviscid flow. We extend his analysis by including the effect of thermal pressure, viscosity, and resistivity, and by considering also the low frequency regime. Furthermore, we focus on the non linear dynamics in a subsonic flow, and show the existence of a non linear pressure feedback. For this purpose we use a combination of MHD simulations and analytic arguments, which we apply to a model of Alfv\'en surface that is voluntarily as simplified as possible. 

In Section~2 we describe our set up and our numerical method. Section~3 gives a qualitative description of the simulations, while Sections~4 and 5 are devoted to a quantitative description of the amplification of Alfv\'en waves in two different frequency regimes. The saturation of the Alfv\'en amplification is studied in Section~6. Finally we discuss the astrophysical consequences of our results in Section~7 and conclude in Section~8.

\section{The model}
\subsection{Stationary flow}
\begin{figure}[tbp]
\centering
\plotone{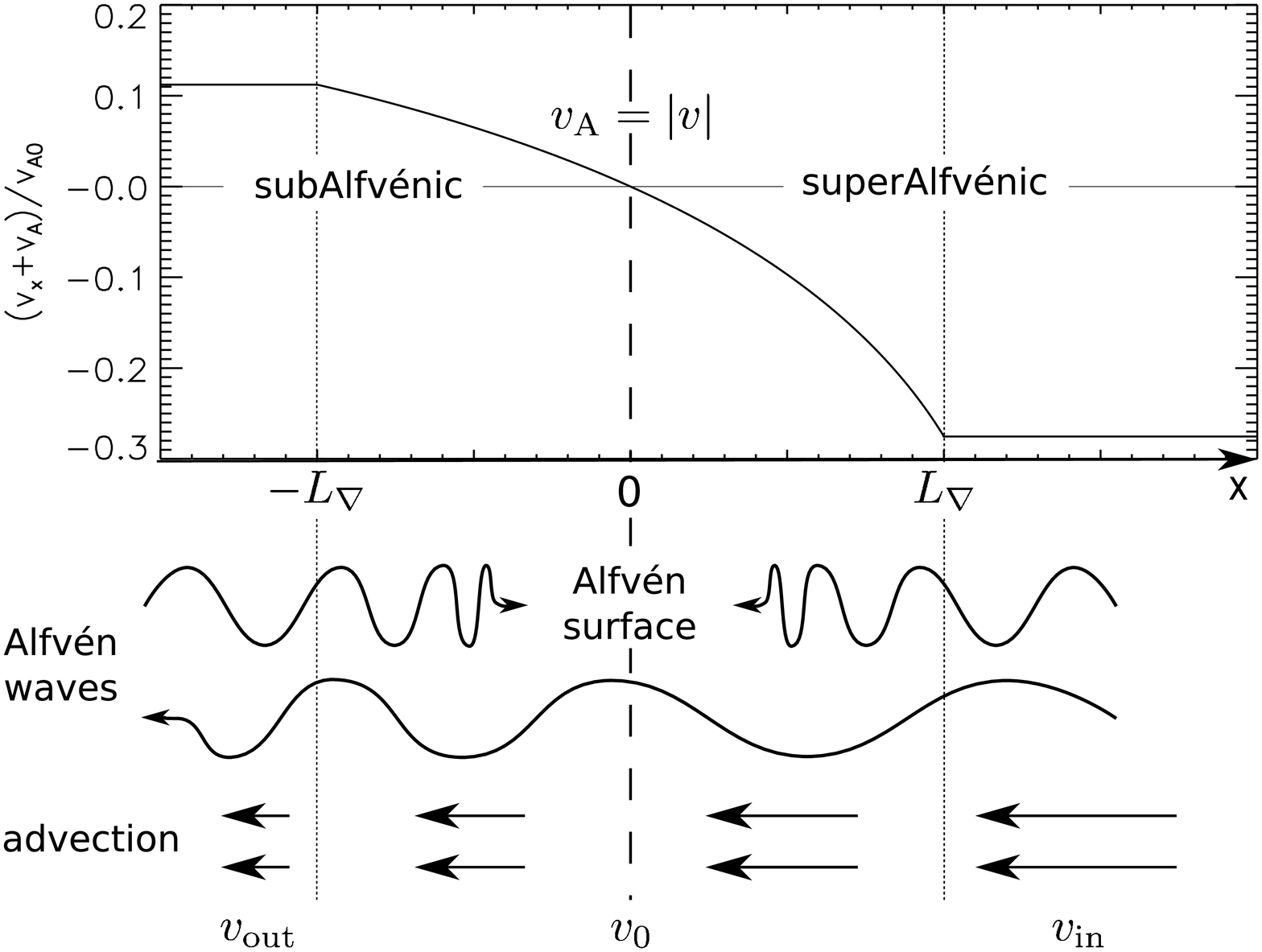}
 \caption{Schematic view of the setup. An external potential step located in $-L_\nabla < x < L_\nabla$ decelerates the flow from a superAlfv\'enic to a subAlfv\'enic velocity. An Alfv\'en wave propagating against the flow is sent from either the lower or the upper boundary, and accumulates at the Alfv\'en surface. The upper part of the figure represents the propagation speed of this wave ($v + v_{\rm A}$), which vanishes at the Alfv\'en surface located in $x=0$. On the contrary, Alfv\'en waves propagating in the same direction as the flow are not affected by the Alfv\'en surface as their propagation velocity $v - v_{\rm A}$ does not vanish.}
             \label{figure1}
\end{figure}

In order to study the dynamics of an Alfv\'en surface in a general perspective, we choose to simplify the physical set up as much as possible. As the phenomena discussed below take place in the vicinity of the Alfv\'en surface, the convergence of the flow is not expected to play a crucial role, hence we restrict our study to a planar geometry. The perfect gas is described by an adiabatic index $\gamma = 4/3$ characterizing a gas of relativistic degenerate fermions. Cooling processes are neglected for the sake of simplicity. The Alfv\'en surface in the stationary flow is created through the adiabatic deceleration of a superAlfv\'enic flow to a subAlfv\'enic velocity by an external potential step (Figure~\ref{figure1}). The superAlfv\'enic flow located at $x>L_\nabla$ has a velocity and a magnetic field that are aligned along the $x$ direction ($v\equiv v_x<0$). The external potential is a linear step extending in the region $-L_\nabla<x<L_\nabla$ and with a depth chosen in such a way that the Alfv\'en surface is located at $x=0$. The incident flow at $x>L_\nabla$ is denoted with the subscript ``in" (e.g. the advection velocity $v_{\rm in}$, sound speed $c_{\rm in}$, and Alfv\'en speed $v_{\rm Ain}$), the downstream flow at $x<-L_\nabla$ with the subscript ``out", and the Alfv\'en surface at $x=0$ with the subscript $0$.
 
The stationary flow profile is determined by the conservation of energy, mass, magnetic flux and entropy:
\begin{eqnarray}
 \frac{\p}{\p x} \left( \frac{ v^{2} }{2} + \frac{c^{2}}{\gamma - 1}  + \phi  \right) &=& 0, \\
\frac{\p}{\p x} \left( \rho v \right) & = &0,  \\
\frac{\p}{\p x} \left(B_{x}\right) & = &0, \\
\frac{\p S }{\p x} & = & 0,
\end{eqnarray}
where the dimensionless entropy $S$ is defined by:
\begin{eqnarray}
S \equiv \frac{1}{\gamma - 1}  \log\left\lbrack  \frac{ P}{P_0} \left(\frac{\rho_0}{ \rho}\right)^{ \gamma}  \right\rbrack.
\end{eqnarray}
As a consequence, the sound speed, density and Alfv\'en speed (defined by $v_{\rm A} \equiv B/\sqrt{\mu_0\rho} $) scale with the advection speed as:
\begin{eqnarray}
 c & \propto & v^{\left(1-\gamma \right)/2}, \\
 \rho & \propto & v^{-1}, \\
  v_{A} & \propto & v^{1/2}.
\end{eqnarray}
The Alfv\'en velocity decreases less than the advection velocity, which allows the superAlfv\'enic fluid to become subAlfv\'enic when compressed by the external potential. The potential step has the following depth:
\begin{equation}
\phi_{\rm in}-\phi_{\rm out} =  v_{0}^2 - v_{\rm in}^2 + \frac{2}{\gamma-1}c_{\rm in}^2\left\lbrack\left(\frac{v_{0}}{v_{\rm in}}\right)^{1-\gamma}-1\right\rbrack 
\end{equation}

The free parameters of the stationary flow are thus $L_\nabla$, $v_0$, $v_{\rm in}$ and $c_0$. Other quantities can be obtained from these (e.g. $v_{\rm Ain} = \sqrt{v_{\rm in}v_{0}}$).

\subsection{Alfv\'en waves}
From the upper or the lower boundary, we send either a circularly or a linearly polarized Alfv\'en wave that is propagating against the flow. The transverse magnetic field and velocity of a circularly polarized wave are described by:
\begin{eqnarray}
\delta {\bf B} & = & \delta B_{\rm in} \left\lbrack \cos\left(k_x x -  \omega t\right) {\bf u_y} + \sin\left(k_x x - \omega t\right) {\bf u_z}  \right\rbrack,   \label{dBalfven} \\
\delta {\bf v} & = & - v_{\rm A}\frac{\delta {\bf B}}{B}, 
\end{eqnarray}
where $ {\bf u_y}$ and $ {\bf u_z}$ are unit vectors along the $y$ and $z$ directions, $\delta B_{\rm in}$ is the amplitude of the Alfv\'en wave (it is noted $\delta B_{\rm out}$ when the Alfv\'en wave is sent from below), $\omega$ its frequency and $k_x$ its wave number:
\begin{equation}
k_x = \frac{\omega}{v_{\rm A}+v}. \label{kalfven}
\end{equation}

In order to generate a symmetric displacement with respect to the initial position, the circularly polarized wave is replaced by a linearly polarized one during the first quarter of a period ($ 0 < t < T_{\rm alf}/4 $). A linearly polarized Alfv\'en wave is obtained from the same set of equations by retaining only the $y$ component of the magnetic field in Equation~(\ref{dBalfven}). 

The circularly polarized Alfv\'en wave is an exact solution of the MHD equations in a uniform flow, in the absence of dissipation, if its amplitude $\delta B$ is constant in time and space. However at the time we start to send the Alfv\'en wave the transition from a zero to a non zero amplitude is not an exact solution, and would create sound waves. To avoid this, we also perturb the density, pressure, and speed of the stationary flow using a second order description of the Alfv\'en wave as is explained in Appendix~\ref{app_alfven}.

\subsection{Parameters and dimensionless numbers}
Unless otherwise specified, the physical quantities are normalized by setting $ v_{A0}=-v_{0}=1$, $\rho_{0}=1$ and $L_{\nabla} = 1$. The free parameters of our setup are then: $v_{\rm in}$, $c_{\rm in}$, the resistivity $\eta$, and the viscosity $\nu$. Alfv\'enic perturbations are characterized by their frequency $\omega$ and their amplitude $\delta B_{\rm in}$ (or $\delta B_{\rm out}$ if sent from below).  We explored the parameter space by  varying the parameters one by one, using the following baseline values: $c_{\rm in} = 2v_0$, $v_{\rm in} = 1.5v_0$, $\eta = 2.10^{-5}$, $\nu = 0$, $\delta B_{\rm in} = 10^{-4}B_0$, and $\omega = 1$. 

We define below a few physical quantities which  are important for the dynamics of the Alfv\'en surface and will be useful in the rest of the paper. The strength of the gradients is described by:
\begin{equation}
\omega_\nabla  \equiv - \frac{\p (v+v_{A})}{\p x} \sim - \frac{v_{\rm Ain}+v_{\rm in}}{L_{\nabla}}.
\end{equation} 
It is also a typical growth rate of the Alfv\'en wave amplitude (Section 4.1). Its ratio with the wave frequency defines a dimensionless number $\epsilon$ that determines whether the wave is in the high frequency regime ($\epsilon \ll 1 $, studied in Section~4) or in the low frequency regime ($\epsilon \gg 1 $, Section~5): 
\begin{equation}
\epsilon \equiv \frac{\omega_\nabla}{\omega} \sim \frac{1}{k_{\rm in}L_\nabla}. 
\label{defepsilon}
\end{equation}
The illustrative simulation described in Section~3 corresponds to $ \epsilon \simeq 0.15  $, and has a behavior typical of the high frequency regime.

The ratio of kinematic viscosity $\nu$  to resistivity $\eta$ (the magnetic Prandtl number) was varied from 0 to infinity, and the same results were found in these two limiting cases. For simplicity, in most of the paper, we consider only the effect of resistivity ; however the same results would be obtained with viscosity instead. As shown in Sections~4 and 5, a typical length scale for the dissipation can be defined as:
\begin{equation}
x_{\rm dis} \equiv \sqrt{\frac{\eta+\nu}{\omega_\nabla}}.
\end{equation}
Its ratio with the lengthscale of the gradients is related to the Reynolds number:
\begin{equation}
\frac{L_\nabla}{x_{\rm dis}} \simeq \sqrt{\frac{L_{\nabla}\left(v_{\rm Ain} + v_{\rm in} \right)}{\eta+\nu}} \sim \sqrt{R_e}.
\end{equation}
In the simulations presented here the dissipation is varied in the range $ R_e \sim 10^3-10^5$ (with $R_e \simeq 10^4$ for our baseline parameters). The core collapse supernovae fluid can be rather viscous inside the neutrinosphere due to neutrino viscosity and is expected to lie in this range of Reynolds number (Section~7). Most other astrophysical fluids would be expected to be much less dissipative.

\subsection{Numerical method}
We use the code RAMSES \citep{teyssier02,fromang06} to solve the usual MHD equations in one dimension along the $x$-direction. Note however that the velocity and magnetic field are three dimensional vectors with non zero components along the $y$ and $z$ directions. RAMSES is a second order Godunov type code that uses the MUSCL-Hancok scheme to evolve the MHD equations. For the calculations presented in this paper, we used the MinMod slope limiter along with the HLLD Riemann solver \citep{miyoshi05}. The computational domain is a one dimensional box of length $4L_\nabla$, comprised between $-1.5L_\nabla$ and $2.5L_\nabla$ when the Alfv\'en wave is sent from above, and between $-2.5L_\nabla$ and $1.5L_\nabla$ when the Alfv\'en wave is sent from below. 

After the stationary flow has settled to its numerical equilibrium, we start sending an Alfv\'en wave from either the upper or the lower boundary. For this purpose we impose in the ghost cells the conditions described by Equations~(\ref{dBalfven})-(\ref{kalfven}) and (\ref{dP_alfven})-(\ref{dvg_alfven}). We use a zero gradient condition for the other boundary, and checked that the reflections caused by such a choice are negligible (see Section~3).

The dissipation is taken into account explicitly with a kinematic viscosity $\nu$ and a resistivity $\eta$. For given dissipation coefficients the resolution is chosen in the range $\Delta x/L_{\nabla} \sim 5.10^{-4}-4.10^{-2}$ so that the numerical dissipation is significantly smaller than the physical one and does not affect the result: this typically corresponds to $\Delta x \lesssim x_{\rm dis}/5$.

\section{Qualitative description of the simulations}
For the purpose of illustration we first focus on our baseline set of parameters and on the case where the Alfv\'en wave is circularly polarized and sent from above.  The frequency such that $ \omega_\nabla/\omega  \simeq 0.15 $ makes it representative of the high frequency regime studied more quantitatively in Section~4. 

\begin{figure}[tbp]
\centering
\plotone{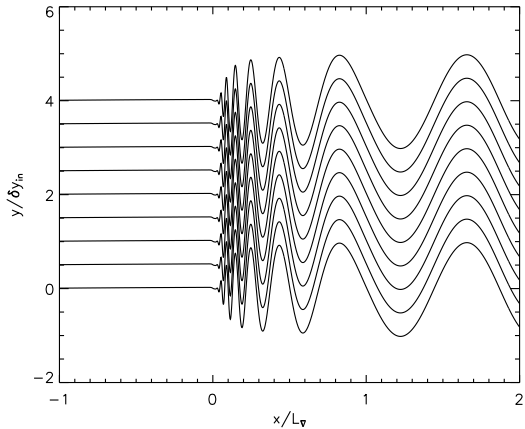}
 \caption{Magnetic field lines projected in the plane (xy) at the end of the simulation with our baseline parameters (a small amplitude circularly polarized incident wave in the WKB regime, $ \omega_\nabla/\omega  \simeq 0.15 $). The scale in the $y$ direction has been normalized  by the displacement of the incident wave $\delta y_{\rm in} = \delta B_{\rm in}/(k_{\rm in}B_0) $. The $x-$axis is normalized by the scale $L_\nabla$ of the stationary gradients. }
             \label{Blines_WKB}
\end{figure}

\begin{figure}[tbp]
\centering
 \plotone{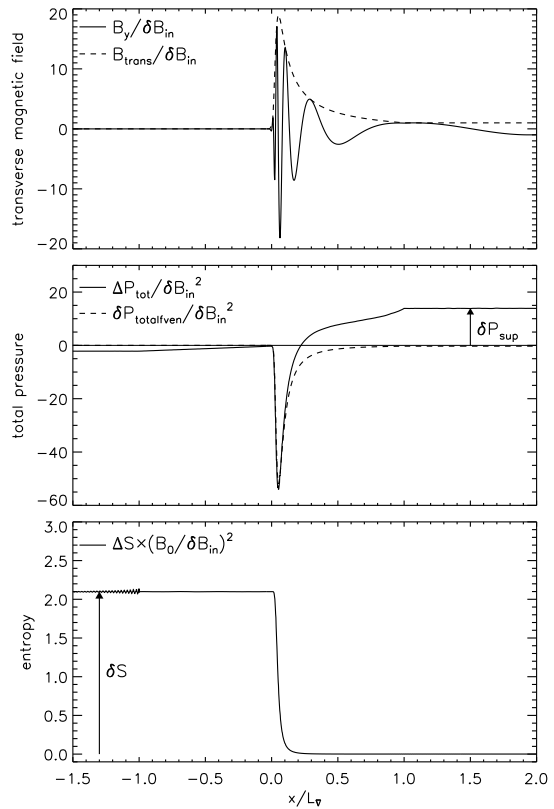}
 \caption{Profiles in the stationary state at the end of the simulation with our baseline parameters. The stationary profiles have been subtracted in order to show the effect of the Alfv\'en wave on the flow. \emph{Upper panel: } $y$ component of the magnetic field (full line) and transverse magnetic field $B_{\rm trans} \equiv \sqrt{B_y^2+B_z^2}$ (dashed line) normalized by the amplitude of the incident Alfv\'en wave. \emph{Middle panel:} total pressure defined as the sum of the thermal pressure and the magnetic pressure $P_{\rm tot} \simeq P + B^2/(2\mu_0)$ (full line). The contribution of the Alfv\'en wave to the total pressure has been estimated using Equation~\ref{dP_alfven} (dashed line). \emph{Bottom panel:} entropy profile. The pressure and entropy have been normalized by $\delta B_{\rm in}^2$ and $\delta B_{\rm in}^2/B_0^2$ in order to define dimensionless quantities that are independent of the incident wave amplitude in the linear regime (see Section~6).  }
             \label{profile_dS_dP}%
\end{figure}

As it approaches the Alfv\'en surface, the Alfv\'en wave is slowed down to a velocity approaching zero, which causes its wavelength to decrease to zero (Figure~\ref{Blines_WKB}). At first, its transverse magnetic field and velocity are amplified to increasingly large values while the transverse displacement stays roughly constant (Figures~\ref{Blines_WKB} and \ref{profile_dS_dP}). The Alfv\'en wave is efficiently damped when the wavelength becomes as small as the dissipative scale, either resistive or viscous. The resulting amplitude profile of the transverse magnetic field peaks close to the Alfv\'en surface, at a maximum value which we refer to as $\delta B_{\rm max}$, and then quickly decreases to zero (Figure~\ref{profile_dS_dP}). This enables the establishment of a stationary state, where the amplitude of the Alfv\'en wave is constant in time (but $\delta B_{yz}$ and $\delta v_{yz}$ are oscillating as the wave is propagating). The dissipation of the Alfv\'en wave is measured at the lower boundary by the change of entropy $\delta S$  as compared to the unperturbed flow (Figure~\ref{profile_dS_dP}, bottom panel). This amplification and dissipation are explained quantitatively in Section~4 for the WKB regime, and in Section~5 for the low frequency regime.

The total pressure profile in Figure~\ref{profile_dS_dP} can be explained by two distinct phenomena. The negative extremum (signifying a decrease as compared to the stationary pressure profile) for $x\gtrsim 0$ is the contribution from the Alfv\'en wave, which amplitude has a similar profile. The second order expansion of the Alfv\'en wave explained in Appendix~\ref{app_alfven} gives a good description of the contribution of the Alfv\'en wave to the total pressure (dashed line in the middle panel of Figure~\ref{profile_dS_dP}). The second effect is the creation of a pressure feedback propagating up and down, which can be visualized by the difference between the dashed line and the full line. The amplitude of the feedback sent upward (respectively downward) is measured at the upper (resp. lower) boundary by $\delta P_{\rm sup}$, $\delta v_{\rm sup}$ and $\delta \rho_{\rm sup}$ (resp. $\delta P_{\rm inf}$, $\delta v_{\rm inf}$ and $\delta \rho_{\rm inf}$). For both signals, we verified that the relation between the pressure, velocity and density perturbations correspond to what is expected from an acoustic wave propagating either up or down (Equations~\ref{eqacoustique1}-\ref{eqacoustique2}). This gives us confidence that the boundary condition we apply does not cause any spurious reflections. We find that the pressure feedback increases the pressure and  decelerates the flow upstream of the Alfv\'en surface, while it decreases the pressure and decelerates the flow downstream. This pressure feedback is explained in Section~4.4.

\begin{figure}[tbp]
\centering
\plotone{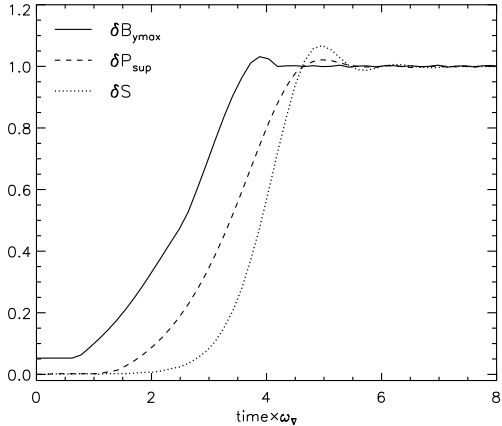}
 \caption{Time evolution in our baseline simulation of the maximum transverse magnetic field $\delta B_{y\rm max}$ (full line), the upstream pressure feedback $\delta P_{\rm sup}$ measured at the upper boundary (dashed line) and the created entropy $\delta S$ measured at the lower boundary (dotted line). $\delta B_{y\rm max}$, $\delta P_{\rm sup}$ and $\delta S$ have been normalized by their final value in the stationary state. The time is normalized using $\omega_{\nabla}$, a typical growth rate for Alfv\'en wave amplification (see Section~4). }
             \label{time_dS_dP}%
\end{figure}

The time evolution of the maximum transverse magnetic field, the pressure feedback, and the entropy creation (shown in Figure~\ref{time_dS_dP}) gives some useful information on the origin of the pressure feedback. The maximum transverse magnetic field increases from the time the Alfv\'en wave enters the gradients at $t\sim 0.5-1$, until $t\sim 4$ when dissipation becomes important. At this time the entropy  starts to increase and quickly reaches its maximum value\footnote{The propagation of entropy and pressure to the boundaries where they are measured induces a delay, which is rather short in both cases: $\Delta t \sim \int^{2.5L_\nabla}_{0}{\rm d}x/(c+v) \sim 0.6\omega_\nabla^{-1}$ for the pressure feedback, and $\Delta t \sim \int^{-1.5L_\nabla}_{0}{\rm d}x/v \sim 0.3 \omega_\nabla^{-1}$ for the entropy.}. On the other hand the pressure feedback increases soon after the beginning of the amplification. This is significantly before the entropy creation starts, indicating that the primary source of this pressure feedback lies in the amplification of the Alfv\'en wave by the gradients rather than in its dissipation.

\section{The WKB regime}
\subsection{Non dissipative waves}
When the Alfv\'en wave has a much shorter wavelength than the typical scale of the gradients, its evolution can be described in the WKB formalism. Close to the Alfv\'en surface, this approximation is valid if the parameter $\epsilon$ defined in Eq.~(\ref{defepsilon}) satisfies $\epsilon \ll 1/4$, which corresponds to the high frequency regime. Note that, in the vicinity of the Alfv\'en surface, this parameter $\epsilon$ does not depend on the distance from the Alfv\'en surface although this may seem counterintuitive as the wavelength decreases to zero \citep{williams75}. This can be understood by the fact that the lengthscale of variation of its effective propagation velocity also scales like the distance $x$ to the Alfv\'en surface.

In the WKB approximation and in the absence of dissipation, the wave action of an Alfv\'en wave is conserved (heuristically this conservation is equivalent to the conservation of wave quanta, see \cite{jacques77}). This property can be used to describe the amplification of the Alfv\'en wave, thus avoiding the lengthy calculations needed for a full WKB approach. The wave action density is defined by:
\begin{equation}
{\cal A} = \frac{E}{\omega_0},    \label{energy_action}
\end{equation}
where  $E$ is the energy density and $\omega_0 = \omega - {\bf k.v}$ is the frequency in the frame of the fluid: 
\begin{eqnarray}
E & = & \frac{1}{2}\left(\rho \delta v^2 + \frac{\delta B^2}{\mu_0}\right) = \frac{\delta B^2}{\mu_0}, \\ 
\omega_0 & = & \omega\frac{v_A}{v_A+v}.
\end{eqnarray}
The conservation of the wave action can then be stated as:
\begin{equation}
 \frac{\p {\cal A}}{\p t} + {\bf \nabla}.\left\lbrack({\bf v+v_{A}}){\cal A}\right\rbrack = 0.
\end{equation}
In a 1D stationary state, the wave action flux is thus constant:
\begin{equation}
(v+v_{A}) {\cal A} = \frac{(v+v_{A})^2}{\omega v_A} \frac{\delta B^2}{\mu_0} = {\rm constant}.
\end{equation}
The following scaling of the Alfv\'en wave amplitude is obtained:
\begin{equation}
\delta B \propto \frac{\sqrt{v_A}}{v_A+v}.
	\label{dB_amplification}
\end{equation}
In the absence of dissipation, the amplitude of an incident Alfv\'en wave thus increases to infinity when approaching the Alfv\'en surface. Such a divergence could be expected from a mere accumulation of the wave energy: assuming a constant energy flux and a vanishing propagation speed implies that the amplitude diverges. What happens to the energy flux of the Alfv\'en wave? The relation between the energy density and the action density (Eq.~\ref{energy_action}) shows that as $\omega_{0}$ is varied, both quantities cannot be conserved at the same time. When an Alfv\'en wave approaches an Alfv\'en surface, the increase of $\omega_0$ implies that the energy flux of the Alfv\'en wave also increases \citep{williams75}:
\begin{equation}
F = (v_A+v)\frac{\delta B^2}{\mu_0} \propto \frac{v_A}{v_A+v}.
\label{energy_amplification}
\end{equation}
This shows that the Alfv\'en wave not only accumulates but actually \emph{amplifies} when approaching the Alfv\'en surface. A wave vector can be associated with this amplification of the energy flux:
\begin{equation}
k_{\rm amp} \equiv \frac{1}{F}\frac{\p F}{\p x} = \frac{1}{v_A} \frac{\p v_{A}}{\p x} - \frac{1}{v+v_{A}} \frac{\p (v+v_{A})}{\p x}.   
\end{equation}
Close to the Alfv\'en surface $v+v_{\rm A} \ll v_{\rm A} $, hence the second term is dominant and may be approximated by:
\begin{equation}
k_{\rm amp} \simeq \frac{\omega_\nabla}{v+v_{\rm A}} \simeq - \frac{1}{x}. 
\label{kamp}
\end{equation}

The growth rate that can be associated with the Alfv\'en wave amplification is:
\begin{equation}
\sigma = k_{\rm amp}(v_{\rm A}+v) \simeq  -\frac{\p (v+v_{A})}{\p x} = \omega_\nabla.
\end{equation}
Thus the steeper the gradient, the faster the amplification. Interestingly, in the idealized case of a compact step, the growth rate would be infinite.

\subsection{Dissipation of Alfv\'en waves by viscosity or resistivity}
In a uniform flow, an Alfv\'en wave obeys the following dispersion relation:
\begin{equation}
k^2v_A^2 = (\omega - {\bf k.v} + i\nu k^2)(\omega - {\bf k.v} + i\eta k^2).
\end{equation}
For a plane wave ($\omega$ real) and small damping rate, the wave vector describing the wave amplitude can be approximated by:
\begin{equation}
k= k_0 + i\frac{\eta + \nu}{2\omega}k_0^3, 
 	\label{kdis}
\end{equation}
where $k_0\equiv\frac{\omega}{v_A+v}$ is the wave vector in the absence of damping. The damping vector describing the energy or the wave action is then twice that of the wave amplitude:
\begin{equation}
k_{\rm dis}= \frac{\eta + \nu}{\omega}k_0^3= \frac{\eta + \nu}{v_{\rm A}+v}k_0^2.
	\label{kdis2}
\end{equation}
In the stationary regime, the wave action is then damped as \citep{jacques77}: 
\begin{equation}
 \frac{\p {\cal }}{\p x} \left\lbrack({\bf v+v_{A}}){\cal A}\right\rbrack = -k_{\rm dis}\left\lbrack({\bf v+v_{A}}){\cal A}\right\rbrack. 
 		\label{action_dis}
\end{equation}
This last equation will be used in Section~\ref{exactWKB}.

\subsection{Maximum amplification and entropy creation}
\subsubsection{Simple estimate}
The maximum energy flux of the Alfv\'en wave can be estimated by stating that amplification and dissipation compensate exactly: $| k_{\rm amp} | = |k_{\rm dis}| $. Using Equations~(\ref{kamp}) and (\ref{kdis}), the wave number $k_{\rm max}$ corresponding to the maximum amplitude can be estimated:
\begin{equation}
k_{\rm max} = \frac{1}{x_{\rm dis}},
\end{equation}
where we implicitly assumed that the maximum amplification happens close to the Alfv\'en surface and  $v+v_{\rm A} \ll v_{\rm A} \simeq v_{\rm A0}$. In order to make a simple estimate, let us assume that the Alfv\'en wave evolves in two steps: the energy flux is amplified without any dissipation in a first step, reaches a maximum $F_{\rm max}$,  and is instantaneously dissipated in a second step. Using Equation~(\ref{energy_amplification}), the maximum energy flux is then:
\begin{eqnarray}
\frac{F_{\rm max}}{F_{\rm in}} & \sim & \frac{v_{A0}}{v_{\rm Ain}}\frac{k_{\rm max}}{k_{\rm in}} =  \frac{v_{A0}}{v_{\rm Ain}}\frac{1}{k_{\rm in}x_{\rm dis}} \sim   \frac{v_{A0}}{v_{\rm Ain}} \epsilon\sqrt{R_{e}}, 
\label{maxflux}
\end{eqnarray}
where the subscript "in" refers to the incident Alfv\'en wave and should be replaced by $out$ in the case where the Alfv\'en wave is sent from below. The entropy created by the dissipation of the amplified Alfv\'en wave can be estimated with this maximum energy flux:
\begin{equation}
\delta S \sim \frac{ F_{\rm max}}{P_0 v_0} \sim  \frac{1}{k_{\rm in}x_{\rm dis}}\frac{F_{\rm in}}{P_0v_{\rm Ain}}. 
	\label{dS_WKB_approx}
\end{equation}
Interestingly the maximum energy flux as well as the created entropy both diverge when the dissipation is decreased to zero, scaling like the square root of the Reynolds number. This suggests a singular behavior and large effects of the Alfv\'en surface in a flow with low dissipation. The nonlinear saturation of the amplification at low dissipation is studied in Section~6.

\subsubsection{Exact calculation \label{exactWKB}}
\begin{figure}[tbp]
\centering
\plotone{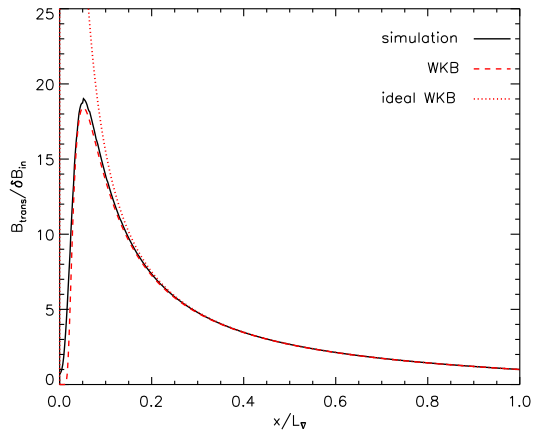}
 \caption{Amplification of an Alfv\'en wave approaching the Alfv\'en point: the transverse magnetic field $\sqrt{B_z^2+B_y^2}$ is shown as function of $x$. The black line is the result of the simulation (at the end of the simulation when a steady state is reached), the dotted line is the prediction by the ideal WKB analysis (Equation~(\ref{dB_amplification})), the dashed line is the WKB prediction that includes the effect of dissipation (deduced from Equation~(\ref{action_profile})). }
             \label{amplification}%
\end{figure}

A more precise calculation can be done in the regime where all the dissipation takes place close to the Alfv\'en surface, so that the velocity profile can be approximated by a linear function:
\begin{equation}
v + v_{\rm A} = -\omega_{\nabla} x.
	\label{vprofile}
\end{equation}
The damping vector can then be written explicitly as a function of $x$ using Equation~(\ref{kdis2}):
\begin{equation}
k_{\rm dis} = - \frac{\omega^2}{\omega_\nabla^2}\frac{x_{\rm dis}^2}{x^3},
\end{equation}
which allows us to solve explicitly Equation~(\ref{action_dis}):
\begin{equation}
({\bf v+v_{A}}){\cal A} =  ({\bf v_{\rm in}+v_{\rm Ain}}){\cal A}_{\rm in}  e^{-\frac{\omega^2}{\omega_\nabla^2}\frac{x_{\rm dis}^2}{2x^2}}.
	\label{action_profile}
\end{equation}
The corresponding transverse magnetic field profile is compared successfully with the simulations in Figure~\ref{amplification} (dashed line versus full line). In particular, the maximum magnetic field is:
\begin{equation}
\delta B_{\rm max} = \sqrt{\frac{2}{e}\frac{v_{\rm A0}}{v_{\rm Ain}}}\frac{\delta B_{\rm in}}{k_{\rm in}x_{\rm dis}}
	\label{dBmax_WKB}
\end{equation}
Using these explicit profiles, the amount of entropy $\delta S$ created can be computed with the ohmic dissipation as:
\begin{equation}
\delta S = \int^{+\infty}_{-\infty} \frac{\eta}{\mu_0 P v}\left( {\bf \nabla} \times {\bf B} \right)^2 \, dx = \sqrt{\frac{\pi}{2}} \frac{1}{k_{\rm in}x_{\rm dis}}\frac{F_{\rm in}}{P_0v_{\rm Ain}},
	\label{dS_WKB}
\end{equation}
which is very close to the simple estimate obtained in Equation~(\ref{dS_WKB_approx}). Note that in the case of a linearly polarized wave, the above equation should be multiplied by a factor $1/2$.

\cite{williams75} called the phenomenon described in this section an instability because the energy of the perturbation (the incoming Alfv\'en wave) increases exponentially. We prefer to call it an amplification, because it critically depends on the properties of the incident Alfv\'en wave. For example the maximum amplitude reached is proportional to its amplitude, and if the source of incident Alfv\'en wave is stopped the perturbations gradually disappear (unless the displacement is kept to a non zero value, see Section~5). Thus contrary to usual instabilities, the growth of perturbations at an Alfv\'en surface needs a continuous incident wave and cannot grow to non linear amplitude from some small initial perturbations: the non linear saturation studied in Section~6 happens only if the incident amplitude is large enough for given dissipation coefficients.

\subsection{Pressure feedback}
In this subsection we investigate the origin of the pressure waves that are emitted up and down when the Alfv\'en wave propagates in the gradients. Some insight about their physical origin can be gained by considering  the energy budget of the flow. As shown in Section~4.1, the energy flux of an Alfv\'en wave in the WKB regime increases when it approaches the Alfv\'en surface, which implies a coupling with at least another wave in order to ensure energy conservation. As the WKB criterion precludes a linear coupling, the energy conservation requires a nonlinear coupling which causes a pressure feedback. Figure~\ref{time_dS_dP} shows that, after a growth period, the pressure feedback levels off and remains at a constant value, suggesting a non linear coupling to a perturbation at zero frequency. By contrast a linear process would have retained the frequency of the incident Alfv\'en wave $\omega$. As will be shown in the following Section, the pressure feedback in the low frequency regime can display a time dependence with a frequency either 0, $\omega$ or $2\omega$. The absence of non zero frequencies in the high frequency regime can be interpreted by the presence of many wavelengths in the gradients, which pressure feedbacks interfere destructively.

In order to estimate quantitatively the pressure feedback, we consider the energy and mass budget of the region $-L_\nabla<x<L_\nabla$. In the stationary state where the pressure feedback is constant, the net energy and mass fluxes should vanish: these two equations are sufficient to determine the two unknowns of the problem (the amplitudes of the acoustic signals propagating up and down). We further assume that the amplification is large, i.e. that the maximum energy flux of the Alfv\'en wave is much larger than the incident energy flux. Neglecting the energy flux due to the incident Alfv\'en wave, the energy and mass flux perturbations are due to: 

(i) the entropy wave, described by:
\begin{eqnarray}
\frac{\delta \rho_{S}}{\rho} & = & \left(1-\gamma\right)\frac{\delta S}{\gamma}, \\
\delta P_{S}  & = & 0, \\
\delta v_{S} & = & 0,
\end{eqnarray}

\begin{figure*}[tbp]
\centering
\plotone{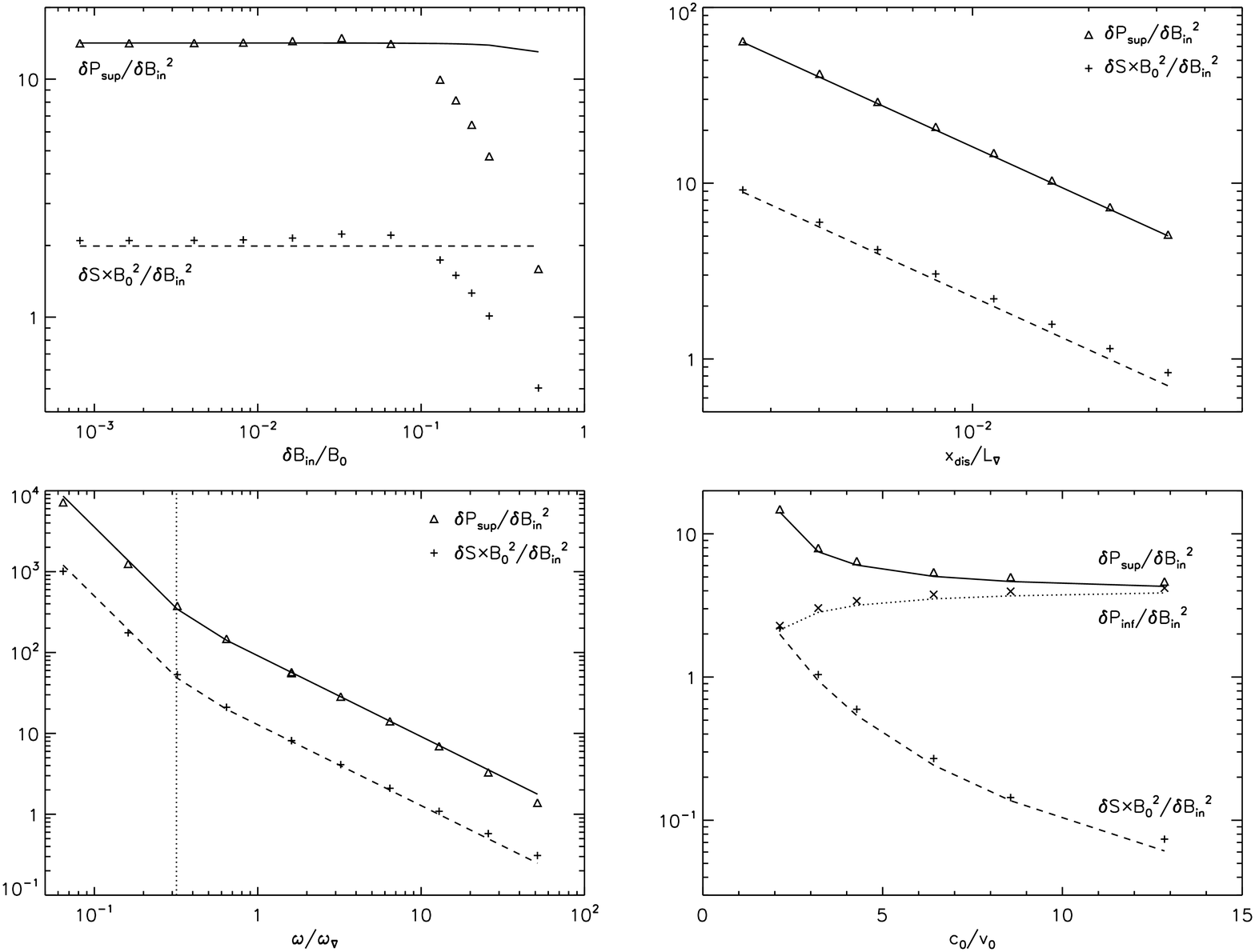}
 \caption{Entropy creation and pressure feedback: comparison between the simulations and the analytical calculation. The four plots show the dependence on the following parameters (from top to bottom and left to right): the amplitude of the incident Alfv\'en wave $\delta B_{\rm in}$, its frequency $\omega$, the resistivity represented here by the parameter $x_{\rm dis} = \sqrt{\eta/\omega_\nabla}$, and the sound speed $c_0$. As each of these parameters is varied, the others are kept constant to their baseline values. The entropy creation $\delta S$ (normalized by $(B_0/B_{\rm in})^2$ in order to be independent of the incident wave amplitude in the linear regime) is shown with crosses for the simulation result, and dashed lines for the analytical prediction. The upstream pressure feedback $\delta P_{\rm sup}$, normalized by the magnetic pressure of the incident wave $\delta B_{\rm in}^2$, is represented by triangles (simulation results) and full lines (analytical prediction). Finally, in the bottom right plot, the pressure feedback downstream is shown by crosses (simulations) and dotted line (analytical) with the same normalization. Notice the two frequency regimes separated by a vertical dotted line at $\omega/\omega_\nabla = 1/\pi$. The agreement between analytics and numerics is excellent, except at large amplitude of the incident wave where non linear effects cause a saturation of the pressure feedback and entropy creation (see Section~\ref{sect_saturation} for more details).}
             \label{couplage}%
\end{figure*}

(ii) the acoustic wave propagating up (termed -) and down (termed +), described by:
\begin{eqnarray}
\delta \rho_\pm & = & \frac{\delta P_\pm}{c^2}, \label{eqacoustique1}\\
 \delta v_\pm & = & \mp \frac{\delta P_\pm}{\rho c}.  \label{eqacoustique2}
\end{eqnarray}
The conservation of mass and energy can be written as follows:
\begin{eqnarray}
(1+\frac{c_{\rm in}}{v_{\rm in}})\frac{\delta P_{-}}{\rho_{\rm in}c_{\rm in}^2} & = & (1 - \frac{c_{\rm out}}{v_{\rm out}})\frac{\delta P_{+}}{\rho_{\rm out}c_{\rm out}^2} + \left(1-\gamma\right)\frac{\delta S}{\gamma}, \\
(1+\frac{v_{\rm in}}{c_{\rm in}})\frac{\delta P_{-}}{\rho_{\rm in}} & = & (1 - \frac{v_{\rm out}}{c_{\rm out}})\frac{\delta P_{+}}{\rho_{\rm out}} + c_{\rm out}^2\frac{\delta S}{\gamma}. 
\end{eqnarray}
Combining these two equations, the pressure perturbation is obtained as a function of the created entropy:
\begin{eqnarray}
\delta P_{\rm sup} = \delta P_{-}  & = & \frac{c_{\rm out}^2-\left(\gamma -1\right)c_{\rm out}v_{\rm out}}{\left(c_{\rm in}+v_{\rm in} \right)\left(c_{\rm in}+\frac{c_{\rm out}v_{\rm out}}{v_{\rm in}} \right)} P_{\rm in}  \delta S,       \label{dP_sup} \\
\delta P_{\rm inf} =  \delta P_{+}  & = &- \frac{c_{\rm out}^2+\left(\gamma -1\right)c_{\rm in}v_{\rm in}}{\left(c_{\rm out}-v_{\rm out} \right)\left(c_{\rm out}+\frac{c_{\rm in}v_{\rm in}}{v_{\rm out}} \right)} P_{\rm out}  \delta S.           \label{dP_inf}
\end{eqnarray}
Using Equation~(\ref{dS_WKB}), the pressure feedback can then be expressed as a function of the incident Alfv\'en wave amplitude and the other parameters of the problem. This analytical result is compared with the simulations in Figure~\ref{couplage}. The dependence on the amplitude and frequency of the incident wave, as well as the resistivity and the Mach number of the flow are reproduced very well. A slight discrepancy appears at very high resistivity because the assumption of large amplification is not fulfilled. For incident waves with a large amplitude, the amplification saturates when the assumption of linear perturbation breaks down: this non linear regime is further studied in Section~6. In addition, a different regime appears at low frequency. This regime is studied in the following section.

\section{The low frequency regime}
\begin{figure*}[tbp]
\centering
\plottwo{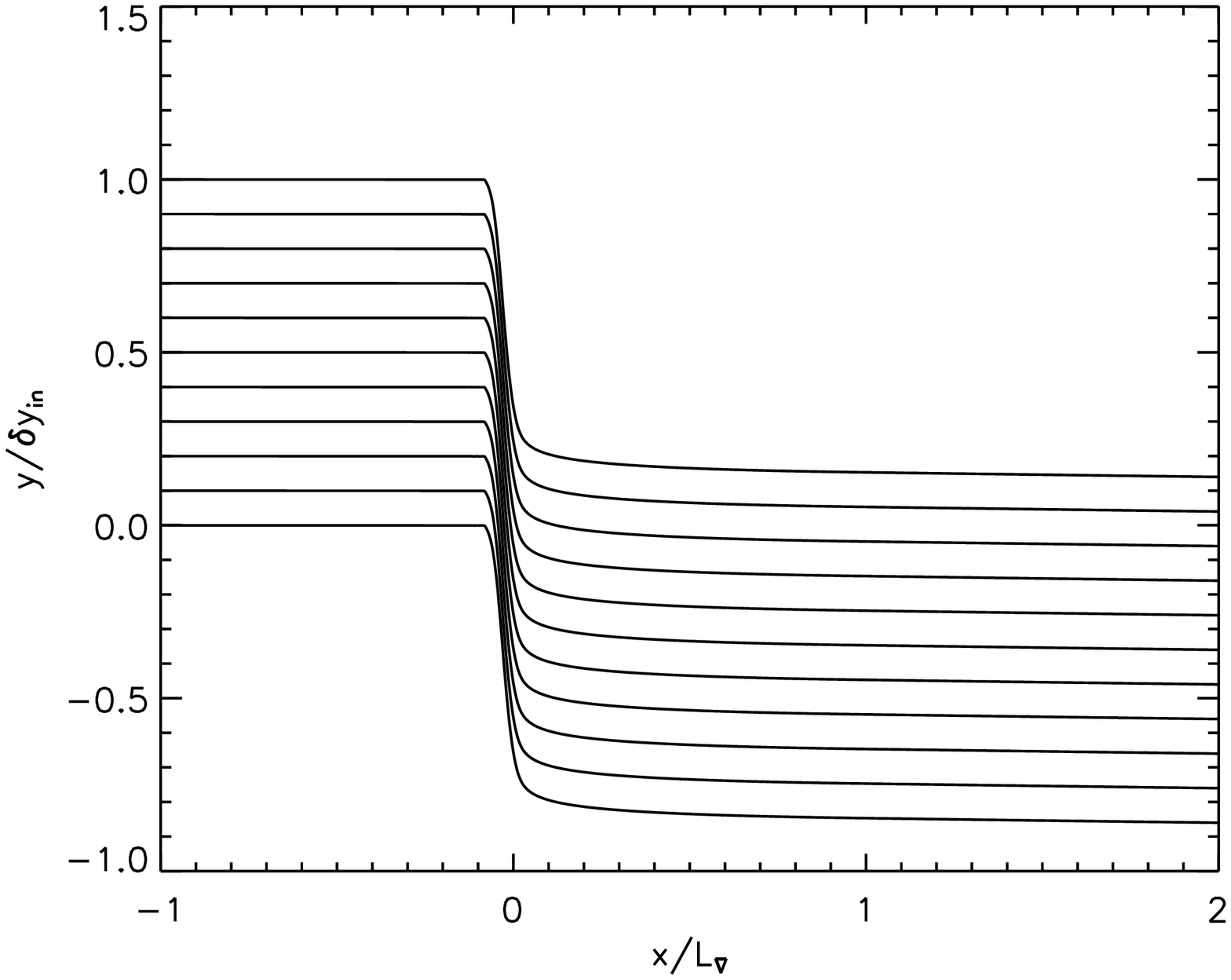}{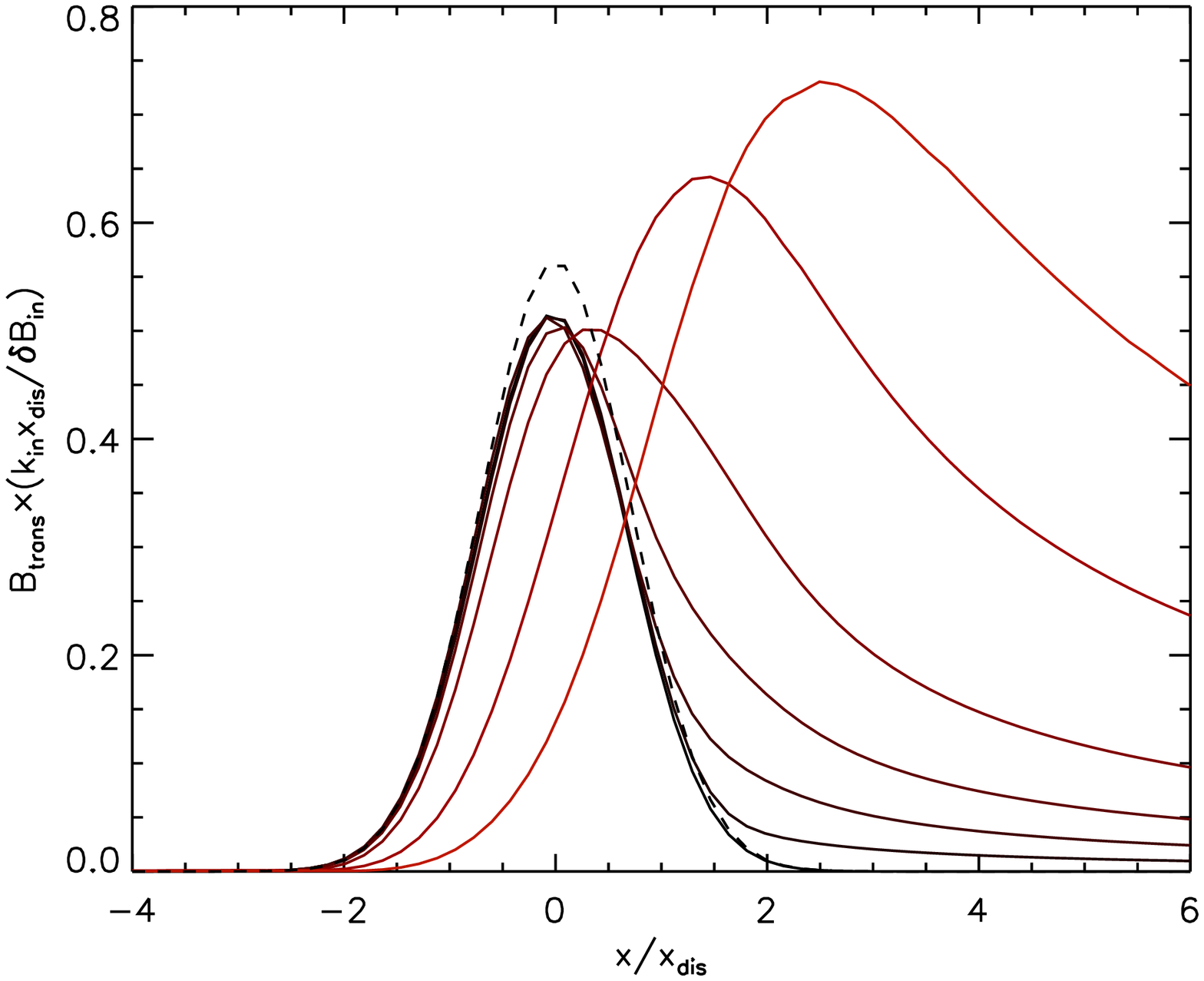}
 \caption{\emph{Left panel:} Magnetic field lines for a circularly polarized Alfv\'en wave in the low frequency regime ($\epsilon \equiv \omega_\nabla/\omega = 15$). \emph{Right panel:} Transverse magnetic field profile (normalized by the displacement amplitude of the incident wave) for different frequency Alfv\'en waves in the low frequency regime compared to the profile produced by a stationary displacement of the field lines. From bottom to top (and black to red in the online version): a stationary displacement, $\epsilon = 15$, $\epsilon = 6$, $\epsilon = 3$, and the three higher curves ($\epsilon = 1.5$, $0.6$ and $0.3$) which are not in the low frequency regime. The dashed line shows the analytic estimate given by Equation~(\ref{gaussian}) and (\ref{Bmax_lowfreq}), which has the correct shape but overestimates the maximum magnetic field by $\sim 10\%$.}
             \label{Blines_lowfreq}%
\end{figure*}

In this section we study the low frequency regime, where the WKB criterion is not satisfied: $\epsilon = \omega_\nabla/\omega \gg 1$. The flow can be considered as stationary at each instant, because the period of the incident Alfv\'en wave is much longer than the response time of the gradients. With this steady state assumption an analytical study of this flow becomes possible in the vicinity of the Alfv\'en surface. As previously, the transverse magnetic field is assumed to be arbitrarily small, and the velocity profile is described by the linear function of Equation~(\ref{vprofile}). As shown in appendix~\ref{app_equadiff}, the transverse magnetic field then follows the second order ordinary differential equation:
\begin{equation}
 \frac{\p^2}{\p x^2}\delta B_y + \frac{2}{x_{\rm dis}^2}\left( x \frac{\p}{\p x}\delta B_y + \delta B_y \right) = 0.
	 \label{equadif_lowfreq}
\end{equation}
Note that the $z$ component follows the same equation. The solution that does not diverge for large $x$ is a Gaussian function with a typical width $x_{\rm dis}$:
\begin{equation}
 \delta B_y =  \delta B_{\rm max} e^{-x^2/x_{\rm dis}^2}.
 	\label{gaussian}
\end{equation}
This suggests that in this regime part of the displacement of the field lines could accumulate in a region of width $x_{\rm dis}$ around the Alfv\'en surface. This is confirmed by the simulations, as shown in Figure~\ref{Blines_lowfreq}. The displacement of the field lines induced by this profile can be computed as:
\begin{equation}
\delta y = \int^{+\infty}_{-\infty} \frac{\delta B_y}{B_0} \, dx = \sqrt{\pi}x_{\rm dis}\frac{\delta B_{\rm max}}{B_0}.
	\label{displacement}
\end{equation}
The entropy created by this profile is:
\begin{equation}
\delta S  =\frac{1}{\sqrt{2\pi}}\frac{\omega_\nabla B_0^2}{\mu_0 P_0v_{0}x_{\rm dis}}\left(\delta y^2+\delta z^2\right).
	\label{dS_displacement}
\end{equation}
The relation between the incident Alfv\'en wave amplitude and $\delta B_{\rm max}$ can then be determined approximately by assuming that the downstream field lines are not perturbed at all. This amounts to neglecting the downward propagating Alfv\'en wave created by linear coupling of the incident Alfv\'en wave in the gradient, and leads to an error of $\sim 10\%$. This simple approximation allows to match the transverse displacement of the incident low frequency Alfv\'en wave with that contained in the vicinity of the Alfv\'en surface (Equation~(\ref{displacement})) and deduce the value of $\delta B_{\rm max}$.  

In order to check the above description, let us first consider a circularly polarized Alfv\'en wave which displacement is symmetric with respect to the initial flow. The displacement of such a wave is constant and equals $ \sqrt{ \delta y^2 + \delta z^2} = \delta B_{\rm in}/(k_{\rm in}B_0)$, giving: 
\begin{equation}
\delta B_{\rm max} = \frac{\delta B_{\rm in}}{\sqrt{\pi}k_{\rm in}x_{\rm dis}}.
	\label{Bmax_lowfreq}
\end{equation}
Figure~\ref{Blines_lowfreq} compares the corresponding profile (dashed line) with the simulations (full lines). This confirms the Gaussian shape and the frequency dependence (contained in the normalization) in the low frequency limit. Furthermore Figure~\ref{Blines_lowfreq} shows that a low frequency Alfv\'en wave behaves like a stationary displacement of the field lines (here induced by sending only half a wavelength of a linearly polarized Alfv\'en wave). As shown in Figure~\ref{Blines_lowfreq} however, Equation~(\ref{Bmax_lowfreq}) overestimates $\delta B_{\rm max}$ by $\sim 10\%$. This can be interpreted by the fact that part of the displacement of the incident wave goes into a downward propagating Alfv\'en wave created by linear coupling in the gradients. A more elaborate  estimate described in Appendix~\ref{app_dBmax} resolves this slight discrepancy. The entropy created by this circularly polarized wave is constant and estimated with Equations~(\ref{dS_displacement}) and (\ref{Bmax_lowfreq}), which can be written in a similar form as Equation~(\ref{dS_WKB}):
\begin{equation}
\delta S =\frac{1}{k_{\rm in}x_\eta} \frac{\omega_\nabla}{\omega} \frac{F_{\rm in}}{\sqrt{2\pi}P_0v_{0}}.
	\label{dS_lowfreq}
\end{equation}
This formula is compared successfully with the simulations in Figure~\ref{couplage} (left bottom panel).

\begin{figure}[tbp]
\centering
 \plotone{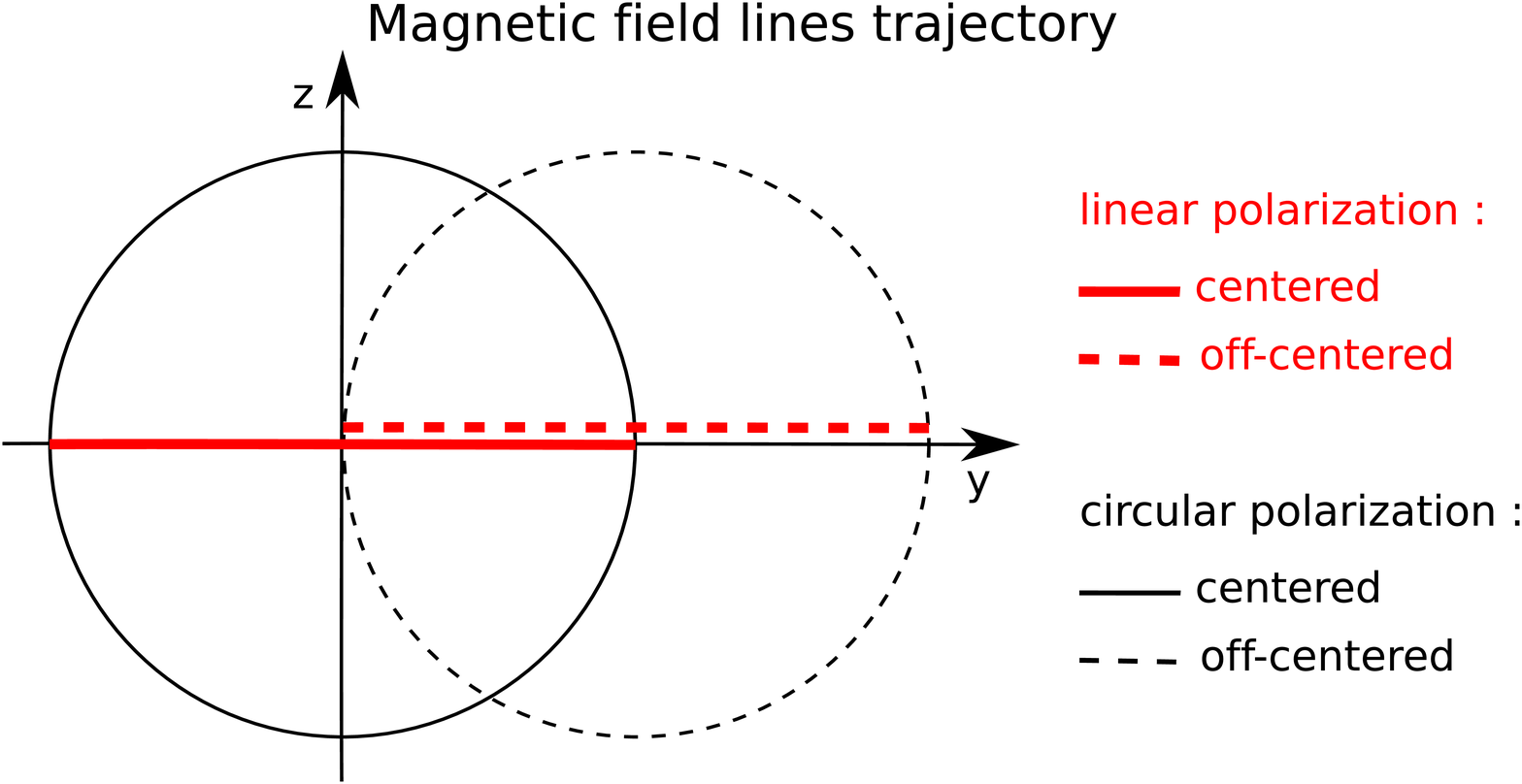}
 \plotone{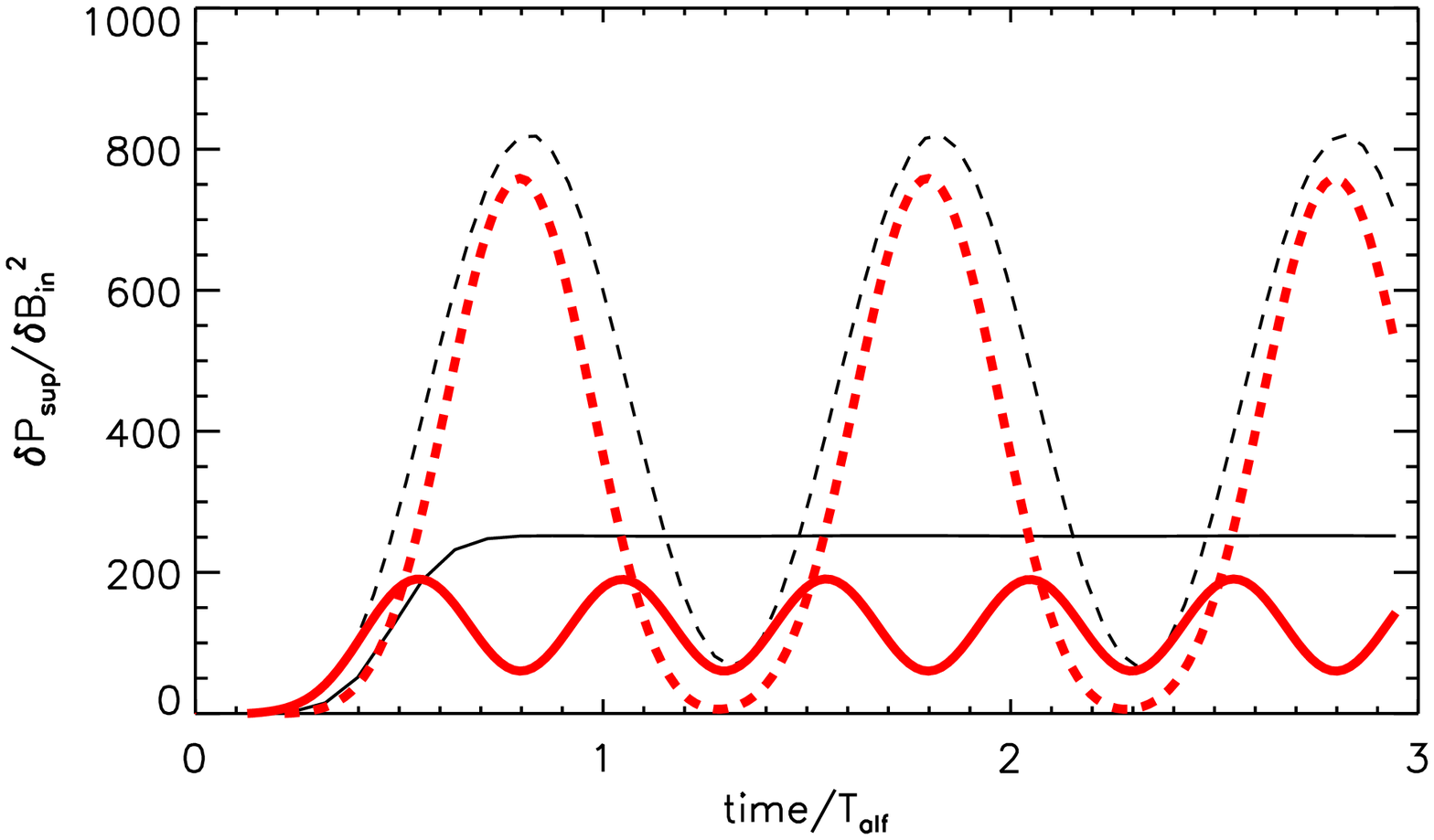}
 \caption{Time dependence of the pressure feedback in the simulations in the low frequency regime (here $\omega=5\times 10^{-2}$, $\epsilon = 3$). The black curves correspond to circularly polarized Alfv\'en waves, the thick gray (red in the online version) curves to linearly polarized waves. The displacement is either symmetric with respect to the initial flow (full lines), or one sided (dashed lines).   }
             \label{circu_lin_time}%
\end{figure}

In the case of other Alfv\'en wave polarizations, the displacement of the field lines (and therefore the entropy created) can be time dependent. However the low frequency assumption ensures that this timescale is much longer than the dynamical timescale, such that the oscillation can be considered as a succession of stationary states and the analysis of Section 4.4 applies. The pressure feedback is proportional to the entropy creation (Equations~(\ref{dP_sup})-(\ref{dP_inf})) and therefore to the square of the displacement. As a consequence the time dependence of the pressure feedback can be understood from the time dependence of the displacement. The trajectory of a field line and the corresponding pressure feedback is illustrated in Figure~\ref{circu_lin_time} in four different cases. For a circularly polarized Alfv\'en wave started as described in Section~2.2, the field line describes a circle around its initial position and thus creates a stationary feedback (i.e. at zero frequency). If the first quarter of a period is omitted (dashed line), it describes an off-centered circle, leading to a displacement and a pressure feedback that oscillates with the same frequency as the incident wave. Similarly, linearly polarized Alfv\'en waves show two different types of time dependence: the centered trajectory leads to an oscillation of the displacement in the range $ - \delta y_0 < \delta y < \delta y_0$ and a pressure feedback with a doubled frequency. The off-centered trajectory oscillates between $0 < \delta y < 2 \delta y_0$, thus the pressure has the same frequency as the incident wave. The maximum displacement of off-centered waves is twice that of centered waves, leading to a maximum pressure feedback that is approximately four times larger (Figure~\ref{circu_lin_time}).

\section{Saturation of the amplification at large amplitude or low dissipation}
	\label{sect_saturation}
When the amplitude of the incident Alfv\'en wave is increased or when the dissipation is decreased, the maximum amplitude of the Alfv\'en wave increases and eventually becomes non linear, and the analysis of the preceding sections no longer applies. This non linear regime  starts when the Alfv\'en wave significantly changes the stationary flow profiles (e.g. of pressure, density, velocity), i.e. when it becomes compressible. In this section we describe this non linear dynamics and show how it affects the Alfv\'en wave amplification and the creation of the pressure feedback. To explain the numerical results, we give an approximate analytical description of the saturation, which takes into account the non linearity of the Alfv\'en wave but not of the pressure feedback. The still higher amplitude regime where the pressure feedback has a significant impact on the flow cannot be studied with our numerical set up, because the acoustic wave interacts with the grid boundaries in a non physical way.

The dynamics of an Alfv\'en wave in the linear regime is very similar in the different configurations: whether it is sent from above or from below gives the same amplification and pressure feedback, the only difference between the circular and linear polarization is the time dependence in the low frequency regime. On the contrary, the non linear dynamics described in this section differs for these configurations, we thus describe them separately in the next subsections.

\subsection{Steepening of linearly polarized Alfv\'en waves}
\begin{figure}[tbp]
\centering
\plotone{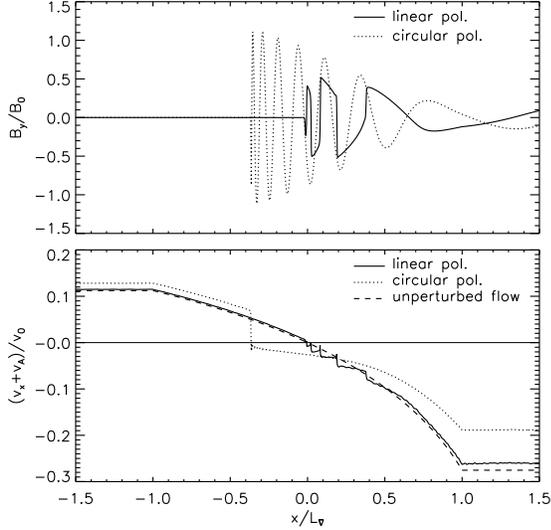}
 \caption{Magnetic field along the $y$ direction (up) and propagation velocity of the Alfv\'en wave vx+vA (bottom) at the end of the simulation. The Alfv\'en waves are sent from above with an amplitude of $\delta B_{\rm in} = 0.13B_0$. The full line corresponds to a linear polarization, the dotted line to a circular polarization, and the dashed line to the unperturbed stationary flow. The linearly polarized wave steepens and therefore dissipates efficiently. On the other hand the circularly polarized wave does not steepen, reaches larger amplitudes and pushes the Alfv\'en surface downward.}
             \label{profile_up}
\end{figure}

The propagation velocity of an Alfv\'en wave non linearly changes as described by $v+v_A$ in Equation~(\ref{dvg_alfven}). Thus the propagation velocity of a maximum of transverse magnetic field pressure is different from that of a minimum, leading to a steepening of the wave (Figure~\ref{profile_up}). The amplitude above which this steepening happens can be estimated with the criterion that the minimum and the maximum (separated by one quarter of a wavelength) have time to catch up before the wave reaches the Alfven surface: 
 \begin{equation}
 \delta^2(v+v_{\rm A})\frac{x}{v+v_{\rm A}} \sim \frac{\lambda}{4} = \frac{\pi}{2}\frac{v+v_A}{\omega}.
 \end{equation}
 Using Equations~(\ref{dB_amplification}) and (\ref{vprofile}), one can get the corresponding transverse magnetic field strength for a given incident amplitude:
 \begin{equation}
 \delta B_{\rm max} \sim  \left(2\pi\frac{\omega_\nabla(v_{\rm in}+v_{\rm Ain})}{\omega\sqrt{v_{\rm A0}v_{\rm Ain}}}\mu_0\rho_0(c_0^2-v_{\rm A0}^2)\delta B_{\rm in}  \right)^{1/3}.
 	\label{dB_steepen}
 \end{equation}
 
\begin{figure}[tbp]
\centering
\plotone{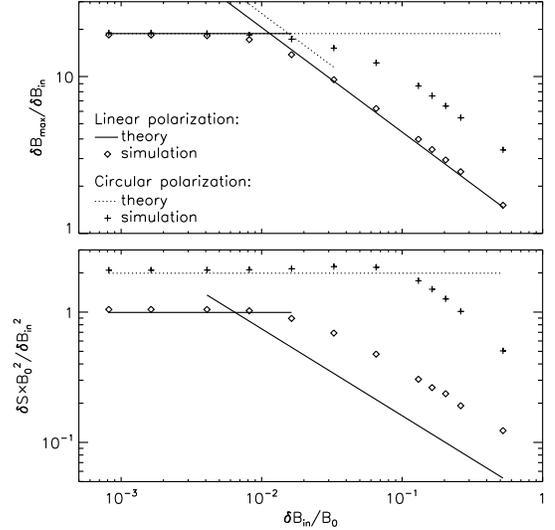}
 \caption{Saturation at large amplitude of linearly versus circularly polarized Alfv\'en waves sent from above. The maximum transverse magnetic field (upper panel) and the entropy created (lower panel) are plotted as a function of the incident wave amplitude. Both quantities are normalized to be constant in the linear regime, so that the onset of the nonlinear regime is clearly visible. Symbols are the simulation results (diamonds for linear polarization, plus signs for circular polarization), while lines represent the analytical estimates (full (dotted) lines for linear (circular) polarization). The two line segments represent the linear solution (horizontal) and nonlinear estimates (note that  for the circular polarization this is only an estimate of the onset of the nonlinear regime, unlike the linear polarization). }
             \label{saturation_circulin}
\end{figure}

The steepening of the wave damps it efficiently, thus Equation~(\ref{dB_steepen}) is a good estimate of the maximum magnetic field reached by a linearly polarized Alfv\'en wave in the non linear regime (Figure~\ref{saturation_circulin}, upper panel). As in Section 4.3.1, we then use the corresponding maximum flux to estimate the amount of entropy created:
 \begin{eqnarray}
 \delta S & \sim & \frac{F_{\rm max}}{P_0v_0}, \\
	&  \sim &  \frac{1}{P_0v_0} \left(2\pi\frac{\omega_\nabla}{\omega}\frac{v_{\rm A0}}{v_{\rm Ain}^2}\mu_0\rho_0(c_0^2-v_{\rm A0}^2) (v_{\rm in}+v_{\rm Ain})^4\delta B_{\rm in}^4  \right)^{1/3}.
 	\label{dS_steepen}
 \end{eqnarray}
This estimate gives the good scaling with $\delta B_{\rm in}$ but is a factor $2$ smaller than the result of the simulations (Figure~\ref{saturation_circulin}, lower panel). This is reasonable given the simplicity of the argument.

\subsection{Circularly polarized Alfv\'en waves}
Circularly polarized Alfv\'en waves do not steepen like the linearly polarized ones, because their magnetic pressure does not oscillate over a wavelength. Even at non linear amplitude, they propagate without deformation in a uniform medium since they are an exact solution of MHD equations. In a non uniform medium however, this is no longer true and  the amplification of the Alfv\'en wave creates a gradient of $v+v_A$ that is non linear (using Equations~(\ref{dvg_alfven}) and (\ref{kamp})):
\begin{equation}
\nabla\left( \delta^2 \left(v+v_A\right) \right) \simeq \omega_\nabla\frac{ v_A}{v+v_{A}}  \frac{ \delta B^2}{2\rho\left( c^2-v_A^2\right)}. 
	\label{nl_gradient}
\end{equation}
This effect is significant if the gradient created by the wave is  comparable to the stationary gradient:
\begin{equation}
\nabla\left( \delta^2 \left(v+v_A\right)\right) \sim \omega_\nabla.
\end{equation}
Using Equations~(\ref{nl_gradient}) and (\ref{dB_amplification}) for a given incident amplitude, the corresponding transverse magnetic field strength is obtained:
 \begin{equation}
 \delta B \sim  \left\lbrack 2\rho_0\left( c_0^2-v_{\rm A0}^2\right)\frac{v_{\rm in}+v_{\rm Ain}} {v_{\rm Ain}}\delta B_{\rm in}\right\rbrack^{1/3}.
 	\label{dB_sat_circu}
\end{equation}
Comparing this with the maximum transverse magnetic field in the linear regime (Equation~(\ref{dBmax_WKB})), gives the threshold of the non linear regime. The behavior is different if the wave is sent from above or from below. The gradient created by an Alfv\'en wave sent from above is opposed to the stationary gradient (because $v+v_A < 0 $ at $x>0$). Therefore it weakens the gradients and pushes the Alfv\'en surface downward, as illustrated in Figure~\ref{profile_up}. The weaker gradients lead to a less efficient amplification and pressure feedback (Figures~\ref{couplage} and \ref{saturation_circulin}).

\begin{figure}[tbp]
\centering
\plotone{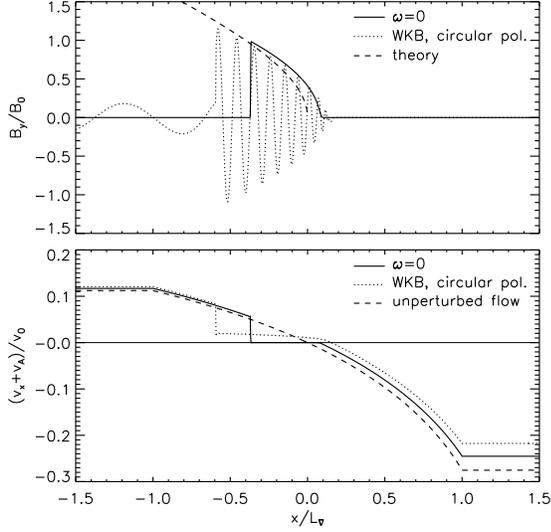}
 \caption{Same as Figure~\ref{profile_up} but for a circular Alfv\'en wave sent from \emph{below} with an amplitude $\delta  B_{\rm out}/B_0 = 0.18 $ (dotted lines), and a stationary displacement of the field lines (full line) which is representative of the low frequency regime. The stationary displacement is induced by sending half a wavelength of a circularly polarized Alfv\'en wave of frequency $\omega = 1$ and amplitude $\delta B_{\rm in}/B_0=0.5$ (the asymmetric wave of Figure~\ref{circu_lin_time}), the result depends only on its displacement $2\delta B_{\rm in}/(B_0k_{\rm in})$.  }
             \label{profile_down}%
\end{figure}

\begin{figure}[tbp]
\centering
\plotone{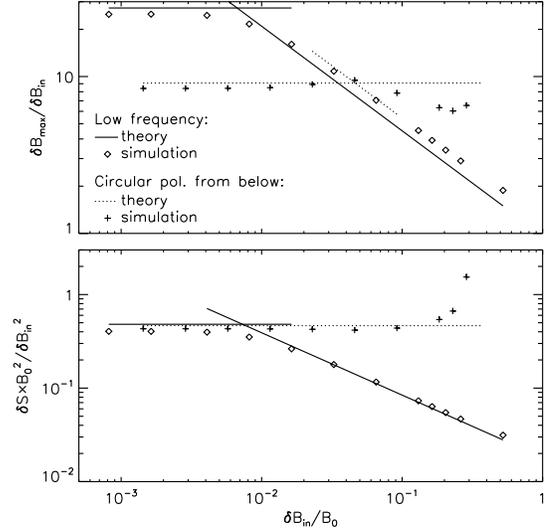}
 \caption{Same as Figure~\ref{saturation_circulin}, but for a circular Alfv\'en wave sent from below (plus signs, dotted lines) and a stationary displacement representative of the low frequency regime (diamonds, full lines). As in Figure~\ref{profile_down} the stationary displacement is induced by sending half a wavelength of a circularly polarized Alfv\'en wave of frequency $\omega = 1$ and amplitude $\delta B_{\rm in}$ (the asymmetric wave of Figure~\ref{circu_lin_time}), the result depends only on its displacement $2\delta B_{\rm in}/(B_0k_{\rm in})$.}
             \label{saturation_baspaquet}%
\end{figure}

In contrast, the gradient induced by a wave sent from below steepens the stationary gradient (because $v+v_A > 0 $ at $x<0$). This causes more amplification which causes more steepening, leading to the build up of a discontinuity (Figure~\ref{profile_down}). Across this discontinuity the flow is accelerated to a velocity close to (though somewhat slower than) the Alfv\'en speed, while the Alfv\'en wave is amplified. Just above the threshold, the discontinuity is weak and close to the Alfv\'en surface. As the amplitude is increased, the discontinuity becomes stronger and moves away from the Alfv\'en surface. Contrary to the case of an Alfv\'en wave sent from above this nonlinear effect does not diminish the amplification, thus no saturation occurs for $\delta B_{\rm in}/B_0<0.3$ (Figure~\ref{saturation_baspaquet}). For larger amplitudes the discontinuity becomes so strong that it reaches the lower grid boundary, leading to non physical effects.

\subsection{The low frequency regime}
As described in Section~5, a low frequency wave in the linear regime accumulates in a region with a width $x_{\rm dis}$ centered on the Alfv\'en surface. In the non linear regime, the magnetic pressure accumulated  around the Alfv\'en surface flattens the velocity profile. Eventually the Alfv\'en wave accumulates over a wider region verifying $v+v_A=0$ and located downstream of the Alfv\'en surface (Figure~\ref{profile_down}). The transverse magnetic field profile can be estimated with the condition $v+v_A=0$:
\begin{equation}
-\omega_\nabla x + \delta^2\left(v+v_A\right) \sim 0,
\end{equation}
thus using Equation~(\ref{dvg_alfven}): 
\begin{equation}
\delta B \sim 2\left\lbrack-\mu_0\rho(c^2-v_A^2)\frac{\omega_\nabla x}{v_A}  \right\rbrack^{1/2}.
	\label{dB_sat_lowfreq}
\end{equation}
This profile is compared with a simulation in Figure~\ref{profile_down}. The (relatively small) discrepancy can be attributed to the effect of the pressure feedback on the velocity profile which was neglected. The width $x_{\rm inf}$ of the region can be approximated as in Section~5 by matching its displacement with that of the incident Alfv\'en wave:
\begin{equation}
x_{\rm inf} \sim \left\lbrack\frac{9}{16}\frac{v_{\rm A0}^3}{\omega_\nabla(c_0^2-v_{A0}^2)}\delta y_{\rm in}^2  \right\rbrack^{1/3},
	\label{xinf_sat_lowfreq}
\end{equation}
thus giving the maximum magnetic field:
\begin{equation}
\delta B_{\rm max} \sim \left\lbrack6\mu_0\rho(c^2-v_A^2)\frac{\omega_\nabla }{v_A}\frac{\delta B_{\rm in}}{k_{\rm in}}  \right\rbrack^{1/3}.
	\label{dB_max_lowfreq}
\end{equation}
The entropy created by this profile can be roughly estimated using:
\begin{equation}
\delta S = \int \frac{\eta}{\mu_0 P v}\left( {\bf \nabla} \times {\bf B} \right)^2 \, dx \sim \frac{\eta\delta B^2}{\mu_0 P v L}, 
\end{equation}
where $L$ is the scale over which $\delta B$ varies. This entropy creation is dominated by the sharp decrease of $\delta B$ located near $-x_{\rm inf}$. The length scale $L$ can be roughly estimated assuming that the width of this sharp decrease is due to the dissipative scale. The width is then such that this diffusion speed due to the resistivity (typically $ v_{\rm diff} \sim \frac{\eta}{L}$) compensates the velocity at which the Alfv\'en wave propagates: $ v_{\rm diff} \sim \omega_\nabla x_{\rm inf} $. Therefore, we get:
\begin{equation}
L \sim \frac{\eta}{\omega_\nabla x_{\rm inf}}.
\end{equation}
Finally, using the four previous equations, the entropy created can be expressed as:
\begin{equation}
\delta S \sim \frac{\rho_0(c_0^2-v_{\rm A0}^2)}{P_0}  \left( \frac{9}{2} \frac{\omega_\nabla^2}{c_0^2-v_{\rm A0}^2} \right)^{2/3} \left(\frac{\delta B_{\rm in}}{Bk_{\rm in}} \right)^{4/3}.
	\label{dS_sat_lowfreq}
\end{equation}

Note that a similar argument in the linear regime gives the same scaling as the more precise calculation of Section~5. The estimates of the maximum magnetic field and entropy creation are compared successfully with the simulations in Figure~\ref{saturation_baspaquet}.

\begin{figure}[tbp]
\centering
\plotone{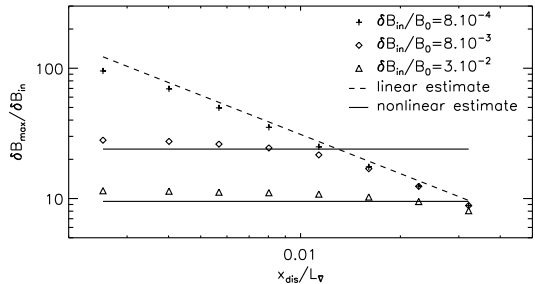}
 \caption{Saturation at low dissipation in the low frequency regime (represented by a stationary displacement, i.e. half a wavelength of a $\omega = 1$ wave). The maximum transverse magnetic field is plotted as a function of the dissipative scale $x_{\rm dis}$, for three values of the incident wave amplitude: $\delta B_{\rm in}/B_0 = 8.10^{-4}$ (plus signs), $\delta B_{\rm in}/B_0 = 8.10^{-3}$ (diamonds) and $\delta B_{\rm in}/B_0 = 3.10^{-2}$ (triangles). The dashed line represents the analytical estimate in the linear regime, while the two full lines correspond to the nonlinear estimate for $\delta B_{\rm in}/B_0 = 8.10^{-3}$  (up) and $\delta B_{\rm in}/B_0 = 3.10^{-2}$ (down).}
             \label{saturation_eta}%
\end{figure}

While the linear amplification diverges at low dissipation, it is important to mention that Equations~(\ref{dB_max_lowfreq}) and (\ref{dS_sat_lowfreq}) show that the nonlinear saturated state is independent of the dissipation coefficient. This is checked with the simulations in Figure~\ref{saturation_eta}. The same phenomenon was found in Section~6.1 for the saturation of linearly polarized waves in the WKB regime (Equations~(\ref{dB_steepen}) and (\ref{dS_steepen})).  This can be interpreted by the fact that the amplitude at which the Alfv\'en wave is dissipated is controlled by nonlinear effects rather than by the dissipative scale.

\section{Consequences for core collapse supernovae}
We now consider the consequences of an Alfv\'en surface in the context of core collapse supernovae, focusing on the phase of stalled accretion shock when matter from the outer layers of the core accretes on the forming proto-neutron star. What maximum amplitude do the Alfv\'en waves reach? Can the pressure feedback significantly affect the shock dynamics? Using the analytic formula derived in this paper, we estimate the conditions under which these effects could be significant.

 \subsection{Origin of the Alfv\'en waves:}
Several instabilities cause non radial motions, which can potentially be a source of Alfv\'en waves: both neutrino driven convection and the standing accretion shock instability (SASI) operate between the neutrinosphere and the shock, while Ledoux convection takes place below the neutrinosphere. Powerful Alfv\'en waves could also be created by the g-modes of the PNS if these are excited to large amplitudes as suggested by \cite{burrows06}. Below we estimate the typical amplitude and frequency of the Alfv\'en waves created by these three phenomena. 

SASI in a magnetized flow was studied linearly by \cite{guilet10a}.  Although SASI can be stabilized by a radial field, this stabilization effect is expected to be less acute if the field is oblique, since the horizontal component can enhance the growth of SASI. \cite{guilet10a} showed that the vorticity created by the shock oscillations is distributed into slow and Alfv\'en waves. The proportion of these two types of waves depends on the field geometry and the wave vector of the SASI mode. Except for very particular cases (for example in a purely radial magnetic field, only slow waves are created) one can expect that a significant fraction of the vorticity is in the form of Alfv\'en waves. Furthermore, downward and upward propagating Alfv\'en waves are created with roughly the same amplitude (strictly in the weak field limit). As a first estimate we will thus assume that the amplitude of the upward propagating Alfv\'en wave of interest for the Alfv\'en surface is a few times smaller than the maximum transverse velocities induced by SASI. According to published numerical simulations, the deformations of the shock induce transverse motions with a speed comparable to or even greater than the sound speed \citep{blondin03,scheck08,marek09}. It is thus reasonable to take an amplitude of $ \delta v \sim 10^4 \,{\rm km.s^{-1}} $, which corresponds to $\delta v/c \sim 0.3$. The oscillation period of SASI is determined by the advection time from the shock to a radius of strong deceleration slightly above the neutrinosphere \citep{foglizzo07,scheck08}, and is usually in the range: $T_{\rm Alf} \sim 30-50 \, {\rm ms }$ ($\omega \sim 100-200\,{\rm s^{-1}}$).

The production of Alfv\'en waves by Ledoux convection in the proto-neutron star was considered by \cite{suzuki08}. The wave amplitude may be estimated by the typical velocities in the PNS convection: $\delta v \sim 3\times 10^3 \, {\rm km.s^{-1}}$ \citep{keil96,buras06b,dessart06}.  The period may be roughly estimated with the overturn time of a convection cell with a size $ l \sim 10 \, {\rm km} $:   $ T_{\rm Alf} \sim l/v \sim 3 \,{\rm ms} $ ($\omega \sim 2.10^3\,{\rm s^{-1}}$). 

\cite{burrows06} witnessed powerful g-mode oscillations that had a typical period of $T \sim 3 \,{\rm ms}$ and a displacement amplitude of $\Delta r \sim 3\, {\rm km} $. The Alfv\'en waves created  by these oscillations may be estimated to have similar amplitude and frequency : $\delta v \sim \omega\Delta r \sim 6000 \,{\rm km.s^{-1}} $. However one should note that such large amplitudes are still controversial since they have not yet been confirmed by other groups \citep{marek09}. Furthermore \cite{weinberg08} suggested that non linear coupling with short wavelength modes unresolved in the simulations could saturate this g-mode at a much lower amplitude.
 
These amplitudes in the range $\delta v \sim 0.3-1.10^4\,{\rm km.s^{-1}}$ are a significant fraction of the sound speed $c\sim 3.10^4\,{\rm km}$ ($\delta v/c \sim 0.1-0.3$). Therefore one would expect such waves to be in the nonlinear regime studied in Section~6.

\subsection{Amplification time scale:}
The timescale of the Alfv\'en wave amplification and pressure feedback is controlled by the stiffness of the velocity gradient:
\begin{equation}
\tau = \omega_{\nabla}^{-1} \sim \left(\frac{{\rm d} v}{{\rm d}r}\right)^{-1} \sim \frac{L_\nabla}{v},
\end{equation}
where $L_\nabla$ is the typical size of the gradient and $v$ a typical advection speed. This timescale depends crucially on the radius of the Alfv\'en surface. If the Alfv\'en surface is inside the proto-neutron star, the advection speed due to the contraction of the PNS is very slow and the timescale could become too long to affect the dynamics of the supernova explosion. 

The situation would be more favorable if the Alfv\'en surface lies outside the PNS. For example, \cite{scheck08} showed that the velocity gradient peaks at a radius slightly above the neutrinosphere and reaches values of $100-300\, {\rm s^{-1}}$ (their Figure~14). This translates into a quite fast growth timescale: $\tau \sim 3-10\,{\rm ms}$. This timescale, associated to the advection time across the deceleration layer with a size $L_\nabla \sim 10-20\, {\rm km}$, is significantly shorter than the period of SASI, which corresponds to the advection time from the shock to the PNS surface on a distance of $r_{\rm sh} - r_{PNS} \sim 100\,{\rm km}$. As a consequence, SASI-created Alfv\'en waves would lie marginally in the low frequency regime ($\omega_\nabla/\omega \sim 1-3$), while those originating from the PNS convection would lie in the WKB regime ($ \omega_\nabla/\omega \sim 0.05-0.15 $).

We conclude that the Alfv\'en wave amplification can be faster than other dynamical timescales, if the Alfv\'en surface lies in the strong gradients slightly above the neutrinosphere. In the next subsection, we discuss whether this corresponds to a realistic magnetic field strength.

\subsection{Magnetic field strength and radius of the Alfv\'en surface: }
Because the infall velocity decreases to zero at the center of the proto-neutron star, an Alfv\'en surface should be present in the flow regardless of the magnetic field strength. However, its consequences and their significance for core collapse supernovae depends on the Alfv\'en surface radius. This radius is controlled by the magnetic field strength, which is unfortunately extremely uncertain. A very weak magnetic field would place the Alfv\'en surface close to the center inside the proto-neutron star, while stronger field would move it outward to larger radii where the infall velocity is larger. 

As discussed in the previous subsection, if the Alfv\'en surface is inside the PNS the amplification timescale may be too long to affect the dynamics before an explosion sets in. Therefore we concentrate on the possibility that the Alfv\'en surface lies above the PNS surface. What strength of the magnetic field is required for this to happen? Assuming an advection speed of $ v \sim 1000\,{\rm km.s^{-1}}$ and a density of $ \rho \sim 10^{10}\, {\rm g.cm^{-3}} $, yields a magnetic field at the Alfv\'en surface of $ B \sim 3.5\times 10^{13}\,{\rm G}$. Compared with other supernovae studies, this corresponds to an intermediate strength. Indeed, magnetic models of supernovae explosions often find or assume a much stronger field of $\sim 10^{14}-10^{15}\,{\rm G}$ with a magnetic pressure comparable to the thermal pressure \citep{akiyama03,shibata06,burrows07b,suzuki08,obergaulinger09}, while here the magnetic pressure remains much smaller than the thermal one: $P_{\rm mag}/P \sim \gamma/2\times v_A^2/c^2 \lesssim 10^{-3} $. On the other hand, this magnetic field strength is significantly larger than the extrapolation from the stellar evolution calculation of \cite{heger05}. 

In order to get an idea of the plausibility of such a field, it is instructive to estimate the magnetic field of the resulting neutron star and to compare it with observed values of pulsars and magnetars. Assuming that the magnetic flux is conserved (with either the usual prescription $B\rho^{2/3}$, or $Br^2$ for a radial field) and extrapolating the magnetic field at the Alfv\'en surface to the final density and radius of a cool neutron star ($\rho \sim 3.10^{14}\, {\rm g.cm^{-3}}$, $r \sim 10\, {\rm km}$) yields $ B \sim 10^{15}-10^{16}\,{\rm G}$. While this is much larger than usual pulsars, it corresponds to the range of magnetic field strength typical of magnetars. Therefore, we conclude that the Alfv\'en surface may play an important role in supernovae leaving a magnetar behind, but should not be significant in the more common supernovae forming normal pulsars. One should however keep in mind that the correspondence between the magnetic field of the PNS and that of much older neutron stars is very uncertain, because the magnetic field might decrease during the supernovae explosion and later evolution, as hypothesized by \cite{wheeler02}. If this were the case, magnetic effects would play a role in a larger subset of supernovae. We should also caution that our description of the Alfv\'en surface dynamics applies only if the magnetic field comes from the progenitor star \citep{ferrario06}, and not if created in situ by a dynamo or the tapping of the rotational energy of the PNS \citep{thompson93}.

\cite{suzuki08} also studied the effect of Alfv\'en waves on the dynamics of core collapse supernovae. They assumed a magnetic field of $B\sim 1-3.10^{15}\, {\rm G}$ at the surface of the proto-neutron star ($r=50\, {\rm km}$), which is one to two orders of magnitude larger than considered in this paper. As a consequence, their Alfv\'en surface lies in the supersonic flow above the shock, while we restricted our study to the case where the Alfv\'en surface is in the subsonic part of the flow.

 \subsection{Viscosity:}
In the linear regime, the maximum amplification and the pressure feedback critically depend on the dissipation scale (due to viscosity or resistivity). However, the large amplitude of the incident Alfv\'en wave would probably place it in the non linear regime discussed in Section~6, in which the dynamics is independent of the dissipation coefficient. Therefore the dissipation should not play an important role unless it is much larger than considered in this paper ; we check below that this is not the case.
 
The hot nuclear fluid constituting the proto-neutron star is an extremely good conductor but is rather viscous due to the diffusion of neutrinos (resulting in a huge magnetic Prandtl number  $P_m \simeq \nu/\eta \sim 10^{13}$, e.g. \cite{thompson93}). The dominant diffusion therefore comes from the neutrino viscosity, which can be estimated in the region optically thick to neutrinos using  the diffusion limit (see e.g. \cite{thompson05}): $\nu \sim 10^{11}\, {\rm cm^2.s{-1}}$, slightly below the neutrinosphere where it is highest. This translates into a dissipation scale $x_{\rm dis} \sim \sqrt{\nu/\omega_\nabla} \sim 0.2\,{\rm km}$ and a typical Reynolds number $Re \sim (L_\nabla/x_{\rm dis})^2 \sim 10^4 $  (using $\omega_\nabla \sim 200\,{\rm s^{-1}}$, $L_\nabla \sim 20\,{\rm km}$ as suggested in the previous paragraph). This value is in the range explored in this paper and yields a significant amplification of an Alfv\'en wave before it is dissipated.  

The steep gradients proposed to be a good place for a strong effect of the Alfv\'en surface lie \emph{above} the neutrinosphere. There, the matter is optically thin to neutrinos and this neutrino viscosity is unfortunately not well defined. However, interactions with streaming neutrinos still have a similar damping effect, which was estimated by \cite{thompson05} by a damping rate $\Gamma \sim 30\, {\rm s^{-1}}$. As this is smaller than the growth rate of the amplification, we do not expect the neutrinos to be able to prevent the amplification very efficiently.

\subsection{Amplitude of the pressure feedback: }
In order to get a simpler scaling of the pressure feedback, we assume that at the Alfv\'en surface, the flow is very subsonic, that is: $ v \sim v_A \ll c $. Equation~(\ref{dP_sup}) giving the pressure feedback propagating toward the shock then simplifies into:
\begin{equation}
\frac{\delta P}{P} \sim \frac{c_{\rm out}^2}{c_{\rm in}^2} \delta S \gtrsim \delta S.
\end{equation}

Since Alfv\'en waves created by convection and g-modes in the proto-neutron star lie in the WKB regime, we use the formulation of Section~6.1. Under the same assumptions, Equation~(\ref{dS_steepen}) simplifies into:
\begin{equation}
\frac{\delta P}{P} \sim \delta S \sim \gamma \left(2\pi\frac{\omega_\nabla}{\omega} \right)^{1/3}\left(\frac{v_{\rm in}+v_{\rm Ain}}{\sqrt{v_{\rm Ain}v_0}}\frac{\delta v}{c} \right)^{4/3}.
\end{equation}
This gives $\delta P/P \sim 0.05$ for PNS convection (using the values: $\omega_\nabla/\omega \sim 0.1$, $ \delta v/c \sim 0.1$ and $(v_{\rm in}+v_{\rm Ain})/\sqrt{v_{\rm Ain}v_0} \sim 1$) and $\delta P/P \sim 0.1$ for g-modes (using $\delta v/c \sim 0.2$).

The Alfv\'en waves created by SASI lie marginally in the low frequency regime, therefore we use the nonlinear formulation developed in Section~6.3. Equation~(\ref{dS_sat_lowfreq}) simplifies into:
\begin{equation}
\frac{\delta P}{P} \sim \gamma\left(\frac{9}{2}\right)^{2/3} \left(\frac{\omega_\nabla}{\omega}\frac{v+v_A}{v_A}\frac{\delta v}{c} \right)^{4/3}
\end{equation}
The values $\omega_\nabla/\omega \sim 2$, $(v+v_A)/v_A \sim 1$ and $ \delta v/c \sim 1/3$ give a very significant pressure feedback: $\delta P/P \sim 2 $. The corresponding maximum transverse magnetic field can be estimated with Equation~(\ref{dB_max_lowfreq}):
\begin{equation}
\delta B_{\rm max} \sim \sqrt{\mu_0\rho} \left(6c^2\frac{\omega_\nabla }{\omega}\frac{v+v_A}{v_A}\delta v  \right)^{1/3},
	\label{dB_max_sasi}
\end{equation}
yielding $\delta B_{\rm max} \sim 1.5\times 10^{15}\,{\rm G}$. Of course this very large value of the pressure feedback should not be taken at face value, since our analysis assumes that the pressure feedback does not significantly modify the upstream flow. A self consistent calculation should consider the shock as the source of Alfv\'en waves in order to capture the effect of the pressure feedback on the Alfv\'en wave injection. We also caution that such a large pressure feedback corresponds to maximum values of the transverse magnetic field with a pressure comparable to the thermal pressure, i.e. much greater than the longitudinal field. Whether this can really happen should be checked further in particular with multidimensional simulations, which might show a different dynamics that could saturate the amplification earlier that in 1D.

We conclude that if the Alfv\'en surface lies in the strong gradients above the neutrinosphere, the amplification of the Alfv\'en waves created by SASI generates a significant pressure feedback. SASI is more efficient than the PNS convection and g-modes because its Alfv\'en waves have a larger amplitude and a lower frequency. A linear acoustic feedback is known to exist in the advective-acoustic cycle, which is believed to be the mechanism driving SASI \citep{foglizzo07,scheck08}. By contrast, this new pressure feedback comes from a non linear coupling. As a consequence, it is potentially important only in the non linear phase of SASI. Also the frequency is not necessarily the same as that of SASI, but has Fourrier components at the frequencies $0$, $\omega_{\rm sasi}$ and $2\omega_{\rm sasi}$. More importantly, instead of oscillating between negative and positive values $\delta P$ is here \emph{always positive}. It is thus expected to be always pushing the shock outwards.

\subsection{Energy budget}
For the processes described in this paper to have a significant impact on the dynamics, the Alfv\'en waves should have access to an appropriate source of energy. As argued in Section~7.1, a fraction of the kinetic energy of SASI and PNS convection could be tapped. If this were the only source of energy, these Alfv\'en waves would probably remain subdominant. However, as shown in Section~4, the energy flux of Alfv\'en waves increases as it approaches the Alfv\'en surface. During this amplification they thus extract an additional energy from the stationary flow. The stationary flow is potentially a much greater energy reservoir than the kinetic energy, if the thermal or the gravitational energy can be tapped. An important question is thus whether the Alfv\'en waves can extract a significant fraction of the thermal energy or if it is limited to the kinetic energy. This depends on the nonlinear saturation of their amplification. In Section~6 we found that the amplification stops roughly when the magnetic pressure is comparable to the thermal pressure, which suggests that the energy of the Alfv\'en waves can become comparable to the thermal energy. One should note however that our simulations were limited to a regime of parameters where the kinetic and thermal energies are comparable. Additional simulations are needed to address this point.

\section{Conclusion}
The main findings of this paper may be summarized as follows:
\begin{itemize}
\item A small amplitude Alfv\'en wave propagating toward an Alfv\'en surface is amplified until its wavelength is as small as the dissipative scale $ x_{\rm dis} $. The timescale for this amplification depends on the stiffness of the gradient: $\tau^{-1} = \p (v+v_{A})/\p x  $.
\item The amplification of the Alfv\'en wave creates a pressure feedback that increases the upstream pressure through a non linear coupling. This pressure feedback is stationary for a high frequency incident wave, while for a low frequency wave it can also display a time dependence with either the same frequency or twice the frequency of the incident wave. 
\item The subsequent dissipation of the amplified Alfv\'en wave leads to an entropy creation, which can be related to the efficiency of the pressure feedback: $\delta P/P \sim \delta S$ (in a very subsonic flow).
\item In the linear regime, the smaller the viscosity/resistivity, the more efficient the amplification and the consequent coupling because the dissipation happens later on (i.e. at smaller scale). The relevant quantities (maximum amplitude, pressure feedback, entropy created etc...) scale as $(\eta + \nu)^{-1/2}$, and could thus be very large in a weakly dissipative flow.
\item At large amplitude of the incident wave or at low dissipation, the amplified Alfv\'en wave becomes non linear and affects the background flow. A linearly polarized wave then steepens, which leads to its efficient dissipation and therefore to the saturation of the amplification. In the low frequency regime, the nonlinear saturation is mediated by the flattening of the velocity profile in the vicinity of the Alfv\'en surface. 
\item In this nonlinear saturated regime, the dynamics becomes independent of the dissipation coefficients, because the amplitude at which the Alfv\'en wave is dissipated is controlled by nonlinear effects rather than by the dissipative scale.
\end{itemize}
An application to core collapse supernovae leads to the following conclusions:
\begin{itemize}
\item Powerful Alfv\'en waves can be created by the proto-neutron star convection, the g-mode oscillations and the SASI induced shock oscillations.
\item The amplification of these waves near the Alfv\'en surface can be fast enough to affect the dynamics, if the magnetic field is strong enough to place this surface in the region of strong deceleration just above the neutrinosphere. Assuming magnetic flux conservation, this would correspond to the formation of a magnetar.
\item In this case, the pressure feedback is dominated by the SASI created Alfv\'en waves, and could contribute significantly to the pressure below the shock. The magnetic field could be amplified to a strength of $\sim 10^{15}\,{\rm G}$. The additional pressure can be expected to push the shock to larger radius, which might be favorable for the success of a potential neutrino driven explosion.
\end{itemize}

Our setup is voluntarily chosen as simplified as possible in order to show the physical processes at work in their simplest form and to render analytical work tractable. Because of its simplicity, our analysis suffers from the following major limitations:
\begin{itemize}
\item We restricted our study to a one dimensional geometry. It is possible that multidimensional phenomena neglected here stop the amplification of Alfv\'en waves at a somewhat lower amplitude than in one dimension. Multidimensional processes that could dissipate Alfv\'en waves include phase mixing \citep{heyvaerts83}, nonlinear multi-wave interactions which lead to a turbulent cascade \citep{goldreich95}, and parametric decay instability \citep{goldstein78,delzanna01}.  This last process can in principle also happen in one dimension, although we did not see any evidence for it in our simulations.
\item The magnetic field was assumed to be parallel to the advection velocity, which is a very singular situation (for example, no Alfv\'en discontinuity, even non evolutionary, is possible under these circumstances). This might be expected in a very subAlfv\'enic flow where the field lines are able to force the gas to follow them, however there is no reason why this should also be the case in the superAlfv\'enic flow. 
\item Incident Alfv\'en waves were imposed at the boundary condition, without specifying how they were created. Strictly speaking, SASI cannot be invoked in the oversimplified field geometry assumed, because SASI would create only slow waves but no Alfv\'en waves. Alfv\'en waves would be generated by SASI only if the field were oblique. 
 \item We studied only the behavior of pure sinusoidal waves, which are meant to model SASI induced oscillations. This approach is a crude approximation since the flow is known to be turbulent. The interaction of such a flow with the Alfv\'en surface should be the subject of future multidimensional simulations.
\item We used a simple adiabatic equation of state. Neglecting the cooling by neutrinos renders our stationary pressure profile unrealistic.
\item We assumed a planar geometry. While this is justified in the vicinity of the Alfv\'en surface (where most of the dynamics takes place), the convergence should be taken into account for a self consistent description from the source of Alfv\'en waves to their amplification.
\item We assumed a large scale magnetic field that connects the region of emission and the Alfv\'en surface. These two regions may be magnetically disconnected if the field is turbulent \citep{thompson93,obergaulinger09,endeve10}.
\item Last but not least, we took as a free parameter the strength of the background magnetic field without explaining its origin. We thus implicitly assumed the field to originate from the progenitor star, and our results directly apply only in this scenario. The magnetic field already present in the collapsing core might or might not be sufficiently strong to place the Alfv\'en surface above the neutrinosphere. If it is not strong enough, it could still be amplified in situ by the MRI \citep{obergaulinger09} or by a dynamo process \citep{thompson93}, but the extension of such a field beyond the Alfv\'en surface is uncertain.
\end{itemize}

A realistic description of the Alfv\'en surface dynamics will therefore require further investigation, in particular through multidimensional simulations. One of the remaining questions concerns the behavior of slow waves, which were not present in the set up studied in this paper. In a high beta plasma (in which the sound speed is significantly larger than the Alfv\'en speed) slow waves bear some similarities with Alfv\'en waves, thus they may be expected to behave in a similar way. As their propagation speed is somewhat slower than the Alfv\'en speed, the surface where they would accumulate should be somewhat beneath the Alfv\'en surface. Because their group and phase velocity depend on their transverse structure, we could expect the region where they accumulate to be broader than in the case of Alfv\'en waves.

Based on the fact that the amplification of Alfv\'en waves extracts energy from the stationary flow, \cite{williams75} proposed that  a broad turbulent zone would be created below the Alfv\'en surface. Our analysis suggests instead that the amplification of Alfv\'en waves energy flux creates an acoustic feedback propagating up and down rather than a turbulent zone. In addition to the inability of our one dimensional analysis to describe turbulence, the difference may be attributed to the fact that we study a subsonic Alfv\'en surface while \cite{williams75} studied a cold supersonic plasma, in which no upstream acoustic feedback is possible.

Finally, we note that an Alfv\'en surface could also exist in an accretion flow onto a magnetic object such as T-Tauri stars, magnetic white dwarfs or neutron stars. The dynamics may however be different in these contexts especially if the Alfv\'en surface lies in a supersonic flow. The possibility suggested by \cite{williams75} that turbulence could play a role in these situations deserves further investigations.

\acknowledgments
The authors are grateful to Jean-Jacques Aly, Romain Teyssier, and Christopher Thompson for insightful discussions, as well as to the anonymous referee whose report helped improve the discussion. This work has been partially funded by the Vortexplosion project ANR-06-JCJC-0119.

\appendix

\section{Weakly compressible alfv\'en waves}   
	 \label{app_alfven}
In the linear regime, an Alfv\'en wave perturbs the velocity and the magnetic field perpendicular to the stationary magnetic field as described in Equations~(\ref{dBalfven})-(\ref{kalfven}). On the other hand the perturbations of pressure, density, and longitudinal velocity and magnetic field are zero to first order. This is no longer true in the non linear regime where the Alfv\'en wave becomes compressible. We estimate this  with a second order development, where we take into account terms scaling like $\delta v^2$ and $\delta B^2$ but neglect all higher order terms. The second order perturbation of a variable $X$ is denoted by $\delta^2X$, thus the background stationary flow is perturbed in the following manner : $\rho \simeq \rho + \delta^2 \rho$, $ v \simeq v + \delta^2v$ etc... We then consider the conservation of magnetic flux, entropy, mass and energy at the interface between an unperturbed region and a region containing the Alfv\'en wave. This interface is moving  with respect to the observer frame at the propagation speed of the Alfv\'en wave $v+v_A + \delta^2(v+v_A)$. The conservation of magnetic flux imposes that the longitudinal magnetic field is unperturbed $\delta^2B_x = 0$, and the conservation of entropy yields $\delta^2P = c^2\delta^2\rho$. The conservation of mass and energy across the interface can then be written :
\begin{eqnarray}
\rho\delta^2v - v_A\delta^2\rho & = & 0 \\ 
\left(c^2\frac{\delta^2\rho}{\rho} - v_A\delta^2v\right)\rho v_A + \frac{1}{2}v_A\left(\delta B_y^2 + \delta B_y^2  \right) & = & 0
\end{eqnarray}
Using these relations, one can then get the second order perturbations as a function of the linear wave amplitude $\delta B^2 \equiv \delta B_y^2 + \delta B_z^2 $ :
\begin{eqnarray}
\delta^2 P & = & c^2\delta^2 \rho = - \frac{c^2\delta B^2}{2\mu_0\left( c^2-v_A^2\right)},  \label{dP_alfven} \\
\delta^2 \rho & = &- \frac{\delta B^2}{2\mu_0\left( c^2-v_A^2\right)}, \\
\delta^2  v & = & v_A\frac{\delta^2 \rho}{\rho} = - \frac{v_A\delta B^2}{2\mu_0\rho\left( c^2-v_A^2\right)}, \\
\delta^2 \left(v+v_A\right)& = & \frac{\delta^2  v }{2} = - \frac{v_A\delta B^2}{4\mu_0\rho\left( c^2-v_A^2\right)}.  \label{dvg_alfven}
\end{eqnarray}

\section{The low frequency regime}
\subsection{Differential equation}
	 \label{app_equadiff}
In a stationary state, the induction and the momentum equations are written (here, we use a resistivity and no viscosity however the same result would be found if one did the opposite assumption of neglecting the resistivity):
\begin{eqnarray}
0 &=&{\bf \nabla\times\left(v\times B \right)} + \eta\Delta {\bf B},\\
{\bf \left(v.\nabla\right)v} & = & -\frac{\bf{\nabla}P}{\rho} + \frac{\bf{(\nabla\times B)\times B}}{\mu_0\rho} 
\end{eqnarray}
We linearize the problem by writing:
\begin{eqnarray}
\bf{v} & = & v.\bf{u_x} + \delta \bf{v}, \\
\bf{B} & = & B.\bf{u_x} + \delta \bf{B}, 
\end{eqnarray}
where $v $ and $B_0$ are the velocity and magnetic field in the $x$ direction in the unperturbed flow, while the perturbations $\delta \bf{v}$ and $\delta \bf{B}$ are assumed to be arbitrarily small. Assuming one dimensional perturbations (no dependence on $y$ and $z$) and keeping only the first order perturbations, the $y$ component of the induction and momentum equations can be written:
\begin{eqnarray}
0 & = & B \frac{\p \delta v_y}{\p x} - v\frac{\p \delta B_y }{\p x} - \frac{\p v}{\p x}\delta B_y + \eta \frac{\p^2 \delta B_y }{\p x^2}, \\
v\frac{\p\delta v_y}{\p x} & =& \frac{v_{\rm A}^2}{B}\frac{\p\delta B_y}{\p x}, 
\end{eqnarray}
which can be combined in one second order differential equation:
\begin{equation}
\eta \frac{\p^2 \delta B_y }{\p x^2} + \frac{(v+v_A)(v-v_A)}{v}\frac{\p \delta B_y }{\p x} - \frac{\p v}{\p x}\delta B_y  = 0.
\end{equation}
In the vicinity of the Alfv\'en surface, the profiles can be linearized: $ v \simeq v_0 - 2\omega_\nabla x $, $ v_A \simeq v_{A0} + \omega_\nabla x  $ and $ 2\omega_\nabla x \ll v_0 $. The differential equation then becomes:
\begin{equation}
\eta \frac{\p^2 \delta B_y }{\p x^2} + 2\omega_\nabla\left(x\frac{\p \delta B_y }{\p x} + \delta B_y\right)  = 0,
\end{equation}
which is equivalent to Equation~(\ref{equadif_lowfreq}). The velocity perturbation can be obtained from the magnetic field perturbation as:
\begin{equation}
\delta v_y = -v_{A0}\frac{\delta B_y}{B}
\end{equation}

\subsection{Estimate of the maximum magnetic field}
	 \label{app_dBmax}
\begin{figure}[tbp]
\centering
\includegraphics[width=8cm]{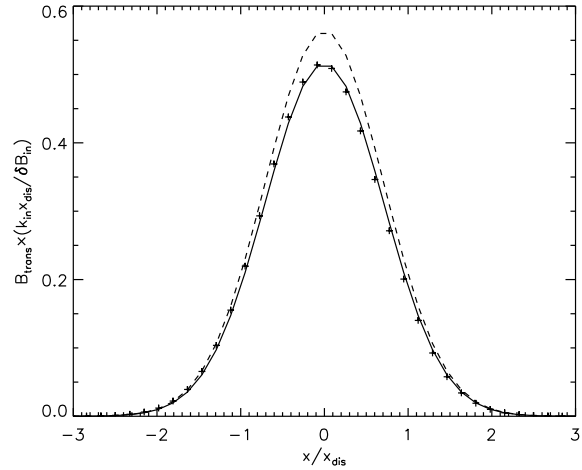}
 \caption{Transverse magnetic profile in the low frequency regime : comparison between the simulation (plus signs), and the analytic estimates. The dashed line is the crude estimate given in Equation~(\ref{Bmax_lowfreq}), while the full line corresponds to Equation~(\ref{Bmax_lowfreq_appendix}). The latter matches the result of the simulations very well.   }
             \label{fig_appendix}%
\end{figure}

In this subsection we estimate better the amplitude of the magnetic field accumulated at the Alfv\'en surface without neglecting the linear coupling with a downward propagating Alfv\'en wave. For this purpose we consider the two following conserved quantities: the transverse momentum, and the displacement of the field line. The downward propagating wave and the displacement accumulated at the Alfv\'en surface should contain the same amount of these quantities as the incident Alfv\'en wave:
\begin{eqnarray}
\delta y_{\rm in} & = & \delta y_0 + \delta y_{\rm out} \\
-\rho_{\rm in}v_{\rm Ain}\delta y_{\rm in} & = & -\rho_{\rm 0}v_{\rm A0}\delta y_{\rm 0} + \rho_{\rm out}v_{\rm Aout}\delta y_{\rm out}
\end{eqnarray}
Finally we get:
\begin{equation}
\delta y_0 = \frac{\rho_{\rm out}v_{\rm Aout} + \rho_{\rm in}v_{\rm Ain}}{\rho_{\rm out}v_{\rm Aout} + \rho_{\rm 0}v_{\rm A0}}\delta y_{\rm in}
\end{equation}
or equivalently:
\begin{equation}
\delta y_0 = \frac{1+ v_{\rm Aout}/v_{\rm Ain}}{1+ v_{\rm Aout}/v_{\rm A0}}\delta y_{\rm in}
	\label{Bmax_lowfreq_appendix}
\end{equation}
This result is compared successfully with the simulations in Figure~\ref{fig_appendix}.

\end{document}